\documentclass[10pt,prd,singlecolumn,aps,nofootinbib,noshowkeys,noshowpacs,superscriptaddress,floatfix]{revtex4-2}
\usepackage{amsmath,amsfonts,amsthm,amssymb}
\usepackage{mathrsfs}
\usepackage{mathtools}
\usepackage[dvips]{graphics,graphicx}
\usepackage[colorlinks=true,linktocpage=true,linkcolor=darkblue,citecolor=red,urlcolor=darkblue]{hyperref}
\usepackage{cancel}
\usepackage{multirow}
\usepackage{longtable}
\usepackage{comment}
\usepackage[normalem]{ulem}
\usepackage{bigints}
\usepackage{xparse}
\usepackage{xcolor}
\usepackage{physics}
\usepackage{verbatim}
\usepackage{tabularx}
\usepackage{minibox}
\usepackage{comment}
\usepackage{appendix}
\usepackage{slashed}
\usepackage{marginnote}
\usepackage[nice]{nicefrac}
 
\usepackage[colorinlistoftodos]{todonotes}
\usepackage{hepunits}
\usepackage{stackengine,scalerel}
\usepackage{tikz}
\usepackage[compat=1.1.0]{tikz-feynman}
\usepackage{pgfplots}
\newcommand\hstar[1]{\ThisStyle{\ensurestackMath{%
\setbox0=\hbox{$\SavedStyle#1$}%
\stackengine{0pt}{\copy0}{\kern.2\ht0\smash{\SavedStyle\star}}{O}{c}{F}{T}{S}}}}
\definecolor{darkblue}{RGB}{0,0,196}
\definecolor{darkgreen}{RGB}{0,120,0}
\definecolor {darkgreen}{rgb}{0.2,0.7,0.2}

\usepackage{float}
\newcommand{\di}{{\rm d}}

\def\spt{{\cal S}}

\newcommand{\be}{\begin{equation}}
\newcommand{\ee}{\end{equation}}                         %
\def\bea{\begin{eqnarray}}
\def\eea{\end{eqnarray}}   
\begin{document}
\title{{Dissipative currents and transport coefficients in relativistic spin hydrodynamics}}
%
%
\author{Asaad Daher}
\email{asaad.daher@ifj.edu.pl}
\affiliation{Institute  of  Nuclear  Physics  Polish  Academy  of  Sciences,  PL-31-342  Krak\'ow,  Poland}
\author{Xin-Li Sheng}
\email{sheng@fi.infn.it}
\affiliation{Università degli studi di Firenze and INFN Sezione di Firenze,\\
Via G. Sansone 1, I-50019 Sesto Fiorentino (Florence), Italy}
\author{David Wagner}
\email{david.wagner@unifi.it}
\affiliation{Università degli studi di Firenze and INFN Sezione di Firenze,\\
Via G. Sansone 1, I-50019 Sesto Fiorentino (Florence), Italy}
\author{Francesco Becattini}
\email{becattini@fi.infn.it}
\affiliation{Università degli studi di Firenze and INFN Sezione di Firenze,\\
Via G. Sansone 1, I-50019 Sesto Fiorentino (Florence), Italy}
%
\begin{abstract}
We determine the form of dissipative currents at the first order in relativistic spin 
hydrodynamics with finite chemical potential including gradients of the spin potential. 
Taking advantage of isotropy in the hydrodynamic local rest frame, using a suitable matching 
condition for the flow velocity and enforcing the semi-positivity of entropy production, we 
find 23 dissipative transport coefficients relating dissipative currents to gradients
of the thermo-hydrodynamic fields: 4 for the symmetric part of the energy-momentum tensor, 
5 for the antisymmetric part, 3 for the conserved vector current, and 11 for the spin tensor.
We compare our finding with previous results in literature.
\end{abstract}
\maketitle
%
\section{Introduction}
%
Relativistic spin hydrodynamics has emerged as a possible theory of phenomenological interest for the
description of the strongly interacting Quark Gluon Plasma in relativistic heavy ion collisions, driven 
by experimental evidence of spin polarization of the produced particles ~\cite{STAR:2017ckg,STAR:2019erd}; 
for a recent review see, for example, Ref.~\cite{Becattini:2024uha}. For this reason, there has been a lot 
of interest in literature on this topic ~\cite{Becattini:2011zz,Florkowski:2017ruc,Hattori:2019ahi,Hattori:2019lfp,Gallegos:2020otk,Li:2020eon,Hongo:2021ona,Fukushima:2020ucl,Gallegos:2021bzp,Montenegro:2017rbu,Montenegro:2018bcf,Bhadury:2020puc,Garbiso:2020puw,Weickgenannt:2020aaf,Shi:2020htn,She:2021lhe,Cartwright:2021qpp,Bhadury:2021oat,Gallegos:2022jow,Peng:2021ago,Liu:2021uhn,Fu:2021pok,Daher:2022xon,Weickgenannt:2022zxs,Hu:2022azy,Daher:2022wzf,Biswas:2023qsw,Daher:2024ixz,Daher:2024bah,Kiamari:2023fbe,Weickgenannt:2023bss,Kumar:2023ojl,Ren:2024pur,Yang:2024duc,Drogosz:2024gzv,Drogosz:2024lkx,Bhadury:2025boe,Florkowski:2024bfw,Dey:2024crk,Dey:2024cwo,Fang:2025aig,Buzzegoli:2024mra,Buzzegoli:2025zud,Tiwari:2024trl},
and yet its physical meaning is still under debate.
In this theory, pseudo-gauge invariance of a combined transformation of energy-momentum and spin tensor 
is broken \cite{Becattini:2018duy}, and, unlike in the most usual formulation of relativistic hydrodynamics, 
the description of the fluid dynamics may need to include a non-vanishing spin tensor if, for instance, the 
system features a separation of scales in the relaxation times of spin and momentum degrees of freedom \cite{Hongo:2021ona}.
However, the relation between the spin tensor and the quantum field operators is not unique; in other
words, the pseudo-gauge fixing remains arbitrary.

For any possible choice of the spin tensor operator, that is of the pseudo-gauge, the conservation equations 
read: 
\begin{align}
\partial_{\mu}T^{\mu\nu}=0~,~~\partial_{\mu}j^{\mu}=0~,~~\partial_{\lambda}J^{\lambda\mu\nu}=0,
\end{align}
where $T^{\mu\nu}$ is the mean value of the energy-momentum tensor, $j^{\mu}$ is the mean value of 
a conserved current, and $J^{\lambda\mu\nu}$ is the mean value of the total angular momentum tensor. 
The latter is decomposed into a so-called orbital part $L^{\lambda\mu\nu}=2T^{\lambda[\nu}x^{\mu]}$,\footnote{We use square (round) brackets for antisymmetrization (symmetrization) of two operators, i.e., $A^{[\mu}B^{\nu]}=(A^\mu B^\nu - A^\nu B^\mu)/2$ and $A^{(\mu}B^{\nu)}=(A^\mu B^\nu + A^\nu B^\mu)/2$.} and a spin tensor:
\begin{align}
J^{\lambda\mu\nu}=L^{\lambda\mu\nu}+S^{\lambda\mu\nu}. 
\end{align}
The conservation of total angular momentum implies:
\begin{align}\label{spincontinuityq}
\partial_{\lambda}S^{\lambda\mu\nu}=T^{\nu\mu}-T^{\mu\nu}.
\end{align}

A key problem in relativistic spin hydrodynamics is the derivation of dissipative 
currents as functions of gradients of thermo-hydrodynamic fields (four-velocity,
temperature, chemical potentials, spin potential or suitable combinations thereof).
Ambiguities may arise between different formulations depending on how these dissipative 
currents are decomposed, which can significantly impact the final physical results. In this 
work, we take advantage of a recent study \cite{Becattini:2023ouz} to provide general 
forms of dissipative currents up to first order in gradients of thermo-hydrodynamic fields. Our method is quite general and provides the maximal number of transport coefficients subject
to the so-called matching condition of vanishing projections of the dissipative part of the currents onto 
the four-velocity field and the condition of positive entropy production. 

The method can be summarized as follows: \textbf{(i)} in general, the dissipative currents are 
expressed as linear combinations of the gradients of thermo-hydrodynamic fields which appear
in the entropy production rate expression, with tensor-valued coefficients;
\textbf{(ii)} we impose that all tensor-valued coefficients are calculated at homogeneous global 
thermodynamic equilibrium such that they are invariant under spatial rotations in the local 
co-moving frame of the fluid. To ensure that this requirement is met, we decompose the tensor 
coefficients in terms of irreducible representations of the rotation group \(\mathrm{SO}(3, \mathbb{R})\) 
and retain only the scalar part.  
\textbf{(iii)} finally, most of the remaining transport coefficients are eliminated by imposing matching 
conditions and the semi-positivity of the entropy production rate. These physical constraints reduce 
the 98 transport coefficients to 23, as tabulated in Sec.~\ref{sec:final-result}, thereby yielding 
the explicit forms of the various dissipative currents.

The paper is organized as follows. In Sec.~\ref{General Tensor Decomposition} we show general expressions
for dissipative currents and how to decompose them with the help of a set of irreducible bases. Then in 
Sec.~\ref{sec:matching-condition} we impose the matching conditions and the semi-positivity constraints, 
which significantly reduces the number of unknown coefficients. The final results for dissipative currents 
and constraints for the corresponding transport coefficients are given in Sec.~\ref{sec:final-result}. 
In Sec.~\ref{QKT}, we verify a subset of these constraints for the specific microscopic framework of quantum kinetic theory. Finally, in Sec.~\ref{Summary} we conclude our results with an outlook.

\section{Dissipative currents and their decomposition} 
\label{General Tensor Decomposition}

In the quantum-statistical description of a relativistic fluid, the entropy production is obtained by calculating 
the divergence of the out-of-equilibrium entropy current, which is constructed from the local equilibrium
density operator $\widehat{\rho}_{\rm LE}$, and the von Neumann entropy 
$S = -\mathrm{Tr}\big[\widehat{\rho}_{\rm LE} \log \widehat{\rho}_{\rm LE}\big]$. The resulting expression 
for the entropy production rate reads~\cite{Becattini:2023ouz},
\begin{align}\label{entropyrate}
\partial_{\mu}s^{\mu}=\left( T_S^{\mu\nu}-T^{\mu\nu}_{S,\,\rm LE}\right) \xi_{\mu\nu}
- \left( j^\mu-j^\mu_{\rm LE}\right) \partial_\mu \zeta  + 
\left( T^{\mu\nu}_{A}-T^{\mu\nu}_{A,\,\rm LE}\right) (\Omega_{\mu\nu}-\varpi_{\mu\nu}) 
- \frac{1}{2}\left( \spt^{\mu\lambda\nu}-\spt^{\mu\lambda\nu}_{\rm LE} \right) \partial_{\mu}\Omega_{\lambda\nu} 
\geq 0\,.
\end{align}
Here, $\xi_{\mu\nu}=\partial_{(\mu}\beta_{\nu)}$ is the thermal shear tensor, 
$\varpi_{\mu\nu}=\partial_{[\nu}\beta_{\mu]}$ is the thermal vorticity tensor, and the fields $\beta_\mu,~\zeta,~\Omega_{\mu\nu}$ denote, respectively, the thermal velocity four-vector $u_\mu/T$, the reduced chemical potential, defined as the chemical potential $\mu$ over temperature $T$, and the reduced spin potential, defined as the spin potential $\omega_{\mu\nu}$ over temperature $T$.

\begin{figure}[htp]
    \centering
    \includegraphics[width=0.4\linewidth]{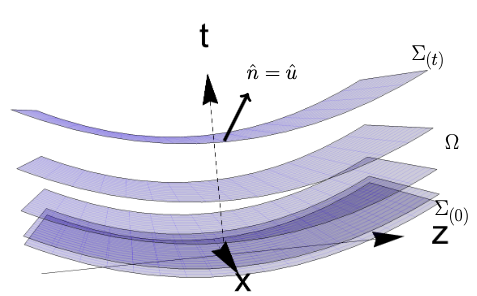}
    \caption{Three-dimensional foliation of initial Cauchy space-like hyperspace $\Sigma_{(0)}$, with volume $\Omega$, and present time hypersurface $\Sigma_{(t)}$, where $\hat{n}$ is chosen to be the fluid four-velocity.}
    \label{fig:1}
\end{figure}

In Eq. \eqref{entropyrate}, quantities with subscript ``LE'’ are mean values of
the quantum operators at local equilibrium, which depend on the definition of a foliation, that
is family of 3D hypersurfaces (see Fig.~\ref{fig:1}). Indeed, local equilibrium values of 
the currents are constrained to provide the exact physical values of the currents projected 
along $n^\mu$, the normal unit vector to the hypersurface, that is:
\begin{align}\label{Matching-conditions}
n_{\mu}(\delta T^{\mu\nu}_{S}+\delta T^{\mu\nu}_{A})=0,~~~n_{\mu}\delta j^{\mu}=0,~~~n_{\mu}\delta 
S^{\mu\lambda\nu}=0\;,
\end{align}
with $\delta X=X-X_\text{LE}$ for any quantity $X$. Provided that $n$ is a vorticity-free vector field (so that orthogonal hypersurfaces may exist)
\begin{equation}\label{vortfree}
\epsilon^{\mu\nu\rho\sigma} n_\nu \partial_\rho n_\sigma = 0\;,
\end{equation}
the choice of $n^\mu$ is arbitrary and corresponds to the freedom of choosing a frame in the
classical formulation of relativistic hydrodynamics. A possible choice for $n^{\mu}$ is the fluid 
4-velocity $u^{\mu}$ (as shown in Fig.~\ref{fig:1}) motivated by the fact that with this choice one 
recovers matching conditions similar to those in Landau's or Eckart's formulations of dissipative hydrodynamics without a spin tensor. However, in the quantum relativistic statistical formulation 
which led to Eq. \eqref{entropyrate}, the condition \eqref{vortfree} applied to the four-velocity
field implies a restriction which is absent in the classical formulation and should be kept in mind when comparing the final results. Particularly, if the velocity field has a finite vorticity (like at global 
thermodynamic equilibrium with non-vanishing angular velocity), $n$ cannot coincide with $u$ and
should be defined otherwise (see the discussion in Ref. \cite{Becattini:2014yxa}).

Equation \eqref{entropyrate} makes it clear which thermo-hydrodynamic gradients contribute to
entropy production. Therefore, we identify these contributions - in the usual classification
scheme - as being of first order in gradients:
\begin{align}
\xi_{\mu\nu}\sim\mathcal{O}(\partial)~,~~\partial_{\mu}\zeta\sim\mathcal{O}(\partial)~,~~(\Omega_{\mu\nu}-\varpi_{\mu\nu})\sim\mathcal{O}(\partial)~,~~\partial_{\mu}\Omega_{\lambda\nu}\sim\mathcal{O}(\partial),
\end{align}
and we will use them to express the dissipative currents at first order in the gradient expansion. The gradient counting order of the term \((\Omega_{\mu\nu} - \omega_{\mu\nu})\) is not assumed but rather imposed by the entropy production, ensuring that all terms in Eq.~\eqref{entropyrate} are of the same gradient order. Consequently, we treat both $\Omega$ and $\varpi$ as quantities that are not small. However, it should be noted that in other studies the reduced spin potential $\Omega$ (and the thermal vorticity as well)
is itself considered as "first-order quantity" in the gradients so that $\partial \Omega$ is neglected
as being a "second order quantity". We do not make such an assumption, thus we tacitly assume $\Omega$
and $\varpi$ to be not necessarily small. On the other hand, we will neglect the anisotropy induced by a potentially large thermal vorticity and spin potential when decomposing the tensor-valued coefficients, as will be made clear shortly.

At first order in gradients we can write in general: 
\begin{align}
&\delta T^{\mu\nu}_{S}=H^{\mu\nu\rho\sigma}\xi_{\rho\sigma}+\frac{1}{T}K^{\mu\nu\rho}\partial_{\rho}\zeta+L^{\mu\nu\rho\sigma}\left(\Omega_{\rho\sigma}-\varpi_{\rho\sigma}\right)+\frac{1}{T}M^{\mu\nu\rho\sigma\tau}\partial_{\rho}\Omega_{\sigma\tau}\label{EMTsymmetricgradientdecomposition}\,,\\
&\delta T^{\mu\nu}_{A}=N^{\mu\nu\rho\sigma}\xi_{\rho\sigma}+\frac{1}{T}P^{\mu\nu\rho}\partial_{\rho}\zeta+Q^{\mu\nu\rho\sigma}\left(\Omega_{\rho\sigma}-\varpi_{\rho\sigma}\right)+\frac{1}{T}R^{\mu\nu\rho\sigma\tau}\partial_{\rho}\Omega_{\sigma\tau}\label{EMTantisymmetricgradientdecomposition}\,,\\
&T\delta j^{\mu}=G^{\mu\rho\sigma}\xi_{\rho\sigma}+\frac{1}{T}I^{\mu\rho}\partial_{\rho}\zeta+O^{\mu\rho\sigma}(\Omega_{\rho\sigma}-\varpi_{\rho\sigma})+\frac{1}{T}F^{\mu\rho\sigma\tau}\partial_{\rho}\Omega_{\sigma\tau}\label{Particlenumbergradientdecomposition}\,,\\
&T\delta S^{\mu\lambda\nu}=T^{\mu\lambda\nu\rho\sigma}\xi_{\rho\sigma}+\frac{1}{T}U^{\mu\lambda\nu\rho}\partial_{\rho}\zeta+V^{\mu\lambda\nu\rho\sigma}\left(\Omega_{\rho\sigma}-\varpi_{\rho\sigma}\right)+\frac{1}{T}W^{\mu\lambda\nu\rho\sigma\tau}\partial_{\rho}\Omega_{\sigma\tau}\label{Spingradientdecomposition}\,,
\end{align}
where we have introduced tensor coefficients $H, K, L, M, \dots, W$ with suitable rank and symmetry in
the exchange of indices; the additional factors $T$ and $1/T$ are introduced such that all the tensor coefficients have the same dimension. 

The tensor coefficients are generally dependent on the quantities that characterize global equilibrium. In general, in Minkowski spacetime, this depends on 10 parameters (plus a set of possible chemical potentials): 
a constant time-like vector and an antisymmetric constant tensor, the thermal vorticity $\varpi$. The possible 
dependencies of the tensor coefficients in \eqref{EMTsymmetricgradientdecomposition}-\eqref{Spingradientdecomposition} 
on those parameters can be, in general, very complicated. In practice, the absence of symmetries of the density 
operator in the most general form of global thermodynamic equilibrium with $\varpi \ne 0$ makes the number of 
relevant scalar coefficients equal to the number of independent components of the tensors. 
However, if we neglect the dependence of those tensor coefficients on thermal vorticity $\varpi$ and we 
expand them from homogeneous global equilibrium, where the four-temperature $\beta$ is constant and $\varpi=0$, 
they must be invariant under spatial rotations in the local co-moving frame of the fluid, and this 
constraint greatly reduces the number of relevant scalar coefficients. 
To fulfill this requirement, we decompose the tensor coefficients in terms of irreducible representations of 
the rotation group ${\rm SO}(3, \mathbb{R})$. The irreducible basis is built on the fluid 4-velocity 
$u^{\mu}$, the transverse projector $\Delta^{\mu\nu}\equiv g^{\mu\nu}-u^\mu u^\nu$, and the rank-3 tensor $\epsilon^{\mu\nu\lambda\gamma}u_\gamma$, 
which are rotationally invariant in the rest frame associated to $u^\mu$. The details on the approach for 
constructing this basis are discussed in Appendix~\ref{AppendixA}.

In Tables \ref{Table-G-I-O-F}-\ref{Table-W}, we list all the irreducible bases and the corresponding factors for each of the tensor coefficients $H^{\mu\nu\rho\sigma}, K^{\mu\nu\rho},\dots, W^{\mu\lambda\nu\rho\sigma\tau}$, where the factors are denoted by lowercase Latin letters with numbers as lower indices. For example,  from Table \ref{Table-G-I-O-F} we can read out the irreducible decomposition of $H^{\mu\nu\rho\sigma}$ as
\begin{equation}
H^{\mu\nu\rho\sigma}=h_1 u^{(\mu}\Delta^{\nu)(\rho}u^{\sigma)}+h_2\Delta^{\mu\nu}\Delta^{\rho\sigma}+h_3 \Delta^{\mu\nu,\rho\sigma}
+h_4\Delta^{\mu\nu}u^{\rho}u^{\sigma}+h_5 u^{\mu}u^{\nu}\Delta^{\rho\sigma}+ h_6u^{\mu}u^{\nu}u^{\rho}u^{\sigma}\;,
\end{equation}
where the traceless symmetric projector $\Delta_{\mu\nu,\alpha\beta}\equiv(\Delta_{\mu\alpha}\Delta_{\nu\beta}+\Delta_{\nu\alpha}\Delta_{\mu\beta})/2-\Delta_{\mu\nu}\Delta_{\alpha\beta}/3$ was used. The tensor coefficients $H^{\mu\nu\rho\sigma}$ (together with $K^{\mu\nu\rho}, \dots, W^{\mu\lambda\nu\rho\sigma\tau}$) are decomposed in equilibrium, which is isotropic in the local co-moving frame of the fluid. In this case, the decompositions do not contain terms involving $\Omega$ because such terms breaks the local rotational invariance. Substituting the decomposed tensor coefficients \( H, K, L, M, \dots, W \) back into the dissipative currents in Eqs.~\eqref{EMTsymmetricgradientdecomposition}-\eqref{Spingradientdecomposition}, and subsequently into the entropy production rate~\eqref{entropyrate}, results in an explicit expression of $\partial_{\mu}s^{\mu}$, which will be analyzed further in the following.

\begin{table}[H]
\centering
\renewcommand{\arraystretch}{1.7}
\begin{tabular}{@{\extracolsep{0.5em}}|c|c||c|c||c|c||c|c||c|c|}
\hline
\multicolumn{2}{|c||}{$H^{\mu\nu\rho\sigma}$} & \multicolumn{2}{c||}{$K^{\mu\nu\rho}$} & \multicolumn{2}{c||}{$L^{\mu\nu\rho\sigma}$} & \multicolumn{4}{c|}{$M^{\mu\nu\rho\sigma\tau}$}\tabularnewline
\hline
$h_{1}$ & $u^{(\mu}\Delta^{\nu)(\rho}u^{\sigma)}$ & $k_{1}$ & $u^{(\mu}\Delta^{\nu)\rho}$ & $l_{1}$ & $u^{(\mu}\Delta^{\nu)[\rho}u^{\sigma]}$ & $m_{1}$ & $u^{(\mu}\Delta^{\nu)[\sigma}u^{\tau]}u^{\rho}$ & $m_{7}$ & $u^{\mu}u^{\nu}u^{[\sigma}\Delta^{\tau]\rho}$\tabularnewline
\hline
$h_{2}$ & $\Delta^{\mu\nu}\Delta^{\rho\sigma}$ & $k_{2}$ & $\Delta^{\mu\nu}u^{\rho}$ & $l_{2}$ & $u^{(\mu}\epsilon^{\nu)\rho\sigma\gamma}u_{\gamma}$ & $m_{2}$ & $u^{(\mu}\epsilon^{\nu)\sigma\tau\gamma}u_{\gamma}u^{\rho}$ & $m_{8}$ & $u^{\mu}u^{\nu}\epsilon^{\sigma\tau\gamma\rho}u_{\gamma}$\tabularnewline
\hline
$h_{3}$ & $\Delta^{\mu\nu,\rho\sigma}$ & $k_{3}$ & $u^{\mu}u^{\nu}u^{\rho}$ &  &  & $m_{3}$ & $\Delta^{\mu\nu}u^{[\sigma}\Delta^{\tau]\rho}$ & $m_{9}$ & $(\epsilon^{\sigma\tau\gamma(\mu}\Delta^{\nu)\rho}+\frac{1}{3}\Delta^{\mu\nu}\epsilon^{\rho\sigma\tau\gamma})u_{\gamma}$\tabularnewline
\hline
$h_{4}$ & $\Delta^{\mu\nu}u^{\rho}u^{\sigma}$ &  &  &  &  & $m_{4}$ & $\Delta^{\mu\nu}\epsilon^{\sigma\tau\gamma\rho}u_{\gamma}$ & $m_{10}$ & $\Delta^{\mu\nu,\rho[\tau}u^{\sigma]}$\tabularnewline
\hline
$h_{5}$ & $u^{\mu}u^{\nu}\Delta^{\rho\sigma}$ &  &  &  &  & $m_{5}$ & $2u^{(\mu}\Delta^{\nu)[\sigma}\Delta^{\tau]\rho}$ &  & \tabularnewline
\hline
$h_{6}$ & $u^{\mu}u^{\nu}u^{\rho}u^{\sigma}$ &  &  &  &  & $m_{6}$ & $u^{(\mu}\epsilon^{\nu)\rho\gamma[\sigma}u^{\tau]}u_{\gamma}$ &  & \tabularnewline
\hline
\end{tabular}

\caption{Irreducible decompositions of $H^{\mu\nu\rho\sigma}$, $K^{\mu\nu\rho}$, $L^{\mu\nu\rho\sigma}$, and $M^{\mu\nu\rho\sigma\tau}$.}
\label{Table-G-I-O-F}

\end{table}

\begin{table}[H]
\centering
\renewcommand{\arraystretch}{1.7}
\begin{tabular}{@{\extracolsep{0.5em}}|c|c||c|c||c|c||c|c||c|c|}
\hline
\multicolumn{2}{|c||}{$N^{\mu\nu\rho\sigma}$} & \multicolumn{2}{c||}{$P^{\mu\nu\rho}$} & \multicolumn{2}{c||}{$Q^{\mu\nu\rho\sigma}$} & \multicolumn{4}{c|}{$R^{\mu\nu\rho\sigma\tau}$}\tabularnewline
\hline
$n_{1}$ & $u^{[\mu}\Delta^{\nu](\rho}u^{\sigma)}$ & $p_{1}$ & $\Delta^{\rho[\mu}u^{\nu]}$ & $q_{1}$ & $u^{[\mu}\Delta^{\nu][\rho}u^{\sigma]}$ & $r_{1}$ & $u^{[\mu}\Delta^{\nu][\sigma}u^{\tau]}u^{\rho}$ & $r_{5}$ & $\epsilon^{\mu\nu\gamma[\sigma}u^{\tau]}u_{\gamma}u^{\rho}$\tabularnewline
\hline
$n_{2}$ & $\epsilon^{\mu\nu\gamma(\rho}u^{\sigma)}u_{\gamma}$ & $p_{2}$ & $\epsilon^{\mu\nu\rho\gamma}u_{\gamma}$ & $q_{2}$ & $2\Delta^{\mu[\rho}\Delta^{\sigma]\nu}$ & $r_{2}$ & $u^{[\mu}\epsilon^{\nu]\sigma\tau\gamma}u_{\gamma}u^{\rho}$ & $r_{6}$ & 2$\Delta^{\mu[\sigma}\Delta^{\tau]\nu}u^{\rho}$\tabularnewline
\hline
 &  &  &  & $q_{3}$ & $u^{[\rho}\epsilon^{\sigma]\mu\nu\gamma}u_{\gamma}$ & $r_{3}$ & 2$u^{[\mu}\Delta^{\nu][\sigma}\Delta^{\tau]\rho}$ & $r_{7}$ & $2\Delta^{\rho[\nu}\epsilon^{\mu]\sigma\tau\gamma}u_{\gamma}$\tabularnewline
\hline
 &  &  &  & $q_{4}$ & $u^{[\mu}\epsilon^{\nu]\rho\sigma\gamma}u_{\gamma}$ & $r_{4}$ & $u^{[\mu}\epsilon^{\nu]\gamma\rho[\sigma}u^{\tau]}u_{\gamma}$ & $r_{8}$ & $2\Delta^{\rho[\mu}\Delta^{\nu][\sigma}u^{\tau]}$\tabularnewline
\hline
\end{tabular}

\caption{Irreducible decompositions of $N^{\mu\nu\rho\sigma}$, $P^{\mu\nu\rho}$, $Q^{\mu\nu\rho\sigma}$, and $R^{\mu\nu\rho\sigma\tau}$.}\label{Table-N-P-Q-R}
\end{table}

\begin{table}[H]
\centering
\renewcommand{\arraystretch}{1.7}
\begin{tabular}{@{\extracolsep{0.5em}}|c|c||c|c||c|c||c|c||c|c|}
\hline
\multicolumn{2}{|c||}{$G^{\mu\rho\sigma}$} & \multicolumn{2}{c||}{$I^{\mu\rho}$} & \multicolumn{2}{c||}{$O^{\mu\rho\sigma}$} & \multicolumn{4}{c|}{$F^{\mu\rho\sigma\tau}$}\tabularnewline
\hline
$g_{1}$ & $\Delta^{\mu(\rho}u^{\sigma)}$ & $i_{1}$ & $\Delta^{\mu\rho}$ & $o_{1}$ & $\Delta^{\mu[\rho}u^{\sigma]}$ & $f_{1}$ & $\Delta^{\mu[\sigma}u^{\tau]}u^{\rho}$ & $f_{4}$ & $\epsilon^{\mu\rho\gamma[\sigma}u^{\tau]}u_{\gamma}$\tabularnewline
\hline
$g_{2}$ & $u^{\mu}\Delta^{\rho\sigma}$ & $i_{2}$ & $u^{\mu}u^{\rho}$ & $o_{2}$ & $\epsilon^{\mu\rho\sigma\gamma}u_{\gamma}$ & $f_{2}$ & $\epsilon^{\mu\sigma\tau\gamma}u^{\rho}u_{\gamma}$ & $f_{5}$ & $u^{\mu}\Delta^{\rho[\sigma}u^{\tau]}$\tabularnewline
\hline
$g_{3}$ & $u^{\mu}u^{\rho}u^{\sigma}$ &  &  &  &  & $f_{3}$ & $2\Delta^{\mu[\sigma}\Delta^{\tau]\rho}$ & $f_{6}$ & $u^{\mu}\epsilon^{\rho\sigma\tau\gamma}u_{\gamma}$\tabularnewline
\hline
\end{tabular}

\caption{Irreducible decompositions of $G^{\mu\rho\sigma}$, $I^{\mu\rho}$, $O^{\mu\rho\sigma}$, and $F^{\mu\rho\sigma\tau}$.}
\end{table}

\begin{table}[H]
\centering\renewcommand{\arraystretch}{1.7}
\begin{tabular}{@{\extracolsep{0.5em}}|c|c||c|c||c|c||c|c|}
\hline
\multicolumn{4}{|c||}{$T^{\mu\lambda\nu\rho\sigma}$} & \multicolumn{2}{c||}{$U^{\mu\lambda\nu\rho}$} & \multicolumn{2}{c|}{$V^{\mu\lambda\nu\rho\sigma}$}\tabularnewline
\hline
$t_{1}$ & $\Delta^{\mu[\nu}u^{\lambda]}u^{\rho}u^{\sigma}$ & $t_{6}$ & $u^{[\lambda}\epsilon^{\nu]\mu\gamma(\rho}u^{\sigma)}u_{\gamma}$ & $u_{1}$ & $\Delta^{\mu[\nu}u^{\lambda]}u^{\rho}$ & $v_{1}$ & $2\Delta^{\mu[\rho}\Delta^{\sigma][\nu}u^{\lambda]}$\tabularnewline
\hline
$t_{2}$ & $\Delta^{\mu[\nu}u^{\lambda]}\Delta^{\rho\sigma}$ & $t_{7}$ & $u^{[\lambda}\Delta^{\nu]\mu,\rho\sigma}$ & $u_{2}$ & $2\Delta^{\mu[\nu}\Delta^{\lambda]\rho}$ & $v_{2}$ & $2\Delta^{\mu[\rho}\epsilon^{\sigma]\nu\lambda\gamma}u_{\gamma}$\tabularnewline
\hline
$t_{3}$ & $2\Delta^{\mu[\nu}\Delta^{\lambda](\rho}u^{\sigma)}$ & $t_{8}$ & $(\Delta^{\mu(\rho}\epsilon^{\sigma)\nu\lambda\gamma}-\Delta^{\rho\sigma}\epsilon^{\mu\nu\lambda\gamma}/3)u_{\gamma}$ & $u_{3}$ & $\epsilon^{\mu\nu\lambda\gamma}u_{\gamma}u^{\rho}$ & $v_{3}$ & $2\Delta^{\mu[\lambda}\Delta^{\nu][\rho}u^{\sigma]}$\tabularnewline
\hline
$t_{4}$ & $\epsilon^{\mu\nu\lambda\gamma}u_{\gamma}u^{\rho}u^{\sigma}$ & $t_{9}$ & $u^{\mu}u^{[\lambda}\Delta^{\nu](\rho}u^{\sigma)}$ & $u_{4}$ & $\epsilon^{\mu\rho\gamma[\nu}u^{\lambda]}u_{\gamma}$ & $v_{4}$ & $u^{[\rho}\epsilon^{\sigma]\mu\gamma[\lambda}u^{\nu]}u_{\gamma}$\tabularnewline
\hline
$t_{5}$ & $\epsilon^{\mu\nu\lambda\gamma}u_{\gamma}\Delta^{\rho\sigma}$ & $t_{10}$ & $u^{\mu}\epsilon^{\lambda\nu\gamma(\rho}u^{\sigma)}u_{\gamma}$ & $u_{5}$ & $u^{\mu}\Delta^{\rho[\nu}u^{\lambda]}$ & $v_{5}$ & $u^{\mu}u^{[\lambda}\Delta^{\nu][\rho}u^{\sigma]}$\tabularnewline
\hline
 &  &  &  & $u_{6}$ & $u^{\mu}\epsilon^{\nu\lambda\rho\gamma}u_{\gamma}$ & $v_{6}$ & $u^{\mu}u^{[\nu}\epsilon^{\lambda]\rho\sigma\gamma}u_{\gamma}$\tabularnewline
\hline
 &  &  &  &  &  & $v_{7}$ & $u^{\mu}\epsilon^{\nu\lambda\gamma[\rho}u^{\sigma]}u_{\gamma}$\tabularnewline
\hline
 &  &  &  &  &  & $v_{8}$ & $u^{\mu}\Delta^{\lambda[\rho}\Delta^{\sigma]\nu}$\tabularnewline
\hline
\end{tabular}

\caption{Irreducible decompositions of $T^{\mu\lambda\nu\rho\sigma}$, $U^{\mu\lambda\nu\rho}$, and $V^{\mu\lambda\nu\rho\sigma}$.}\label{Table-T-U-V}
\end{table}

\begin{table}[H]
\centering
\renewcommand{\arraystretch}{1.7}
\begin{tabular}{@{\extracolsep{0.5em}}|c|c||c|c||c|c|}
\hline
\multicolumn{6}{|c|}{$W^{\mu\lambda\nu\rho\sigma\tau}$}\tabularnewline
\hline
$w_{1}$ & $\Delta^{\mu[\nu}u^{\lambda]}u^{[\sigma}\Delta^{\tau]\rho}$ & $w_{7}$ & $2\Delta^{\mu[\nu}\Delta^{\lambda][\tau}u^{\sigma]}u^{\rho}$ & $w_{13}$ & $u^{\mu}u^{\rho}\Delta^{\lambda[\sigma}\Delta^{\tau]\nu}$\tabularnewline
\hline
$w_{2}$ & $\epsilon^{\mu\lambda\nu\gamma}u_{\gamma}\epsilon^{\rho\sigma\tau\kappa}u_{\kappa}$ & $w_{8}$ & $2\Delta^{\mu[\tau}\epsilon^{\sigma]\lambda\nu\gamma}u_{\gamma}u^{\rho}$ & $w_{14}$ & $u^{\mu}\epsilon^{\lambda\nu\gamma[\sigma}u^{\tau]}u^{\rho}u_{\gamma}$\tabularnewline
\hline
$w_{3}$ & $\Delta^{\mu[\nu}u^{\lambda]}\epsilon^{\sigma\tau\gamma\rho}u_{\gamma}+\epsilon^{\mu\lambda\nu\gamma}u_{\gamma}\Delta^{\rho[\sigma}u^{\tau]}$ & $w_{9}$ & $4\Delta^{\mu[\lambda}\Delta^{\nu][\sigma}\Delta^{\tau]\rho}$ & $w_{15}$ & $u^{\mu}u^{[\nu}\epsilon^{\lambda]\rho\gamma[\sigma}u^{\tau]}u_{\gamma}$\tabularnewline
\hline
$w_{4}$ & $\Delta^{\mu[\nu}u^{\lambda]}\epsilon^{\sigma\tau\gamma\rho}u_{\gamma}-\epsilon^{\mu\lambda\nu\gamma}u_{\gamma}\Delta^{\rho[\sigma}u^{\tau]}$ & $w_{10}$ & $u^{[\lambda}\Delta^{\nu]\mu,\rho\tau}u^{\sigma]}$ & $w_{16}$ & $u^{\mu}\Delta^{\rho[\nu}\Delta^{\lambda][\sigma}u^{\tau]}$\tabularnewline
\hline
$w_{5}$ & $u^{[\nu}\epsilon^{\lambda]\mu\gamma[\sigma}u^{\tau]}u_{\gamma}u^{\rho}$ & $w_{11}$ & $u^{\mu}u^{[\lambda}\Delta^{\nu][\sigma}u^{\tau]}u^{\rho}$ & $w_{17}$ & $u^{\mu}u^{[\nu}\Delta^{\lambda][\tau}\Delta^{\sigma]\rho}$\tabularnewline
\hline
$w_{6}$ & $2\Delta^{\mu[\sigma}\Delta^{\tau][\nu}u^{\lambda]}u^{\rho}$ & $w_{12}$ & $u^{\mu}u^{[\lambda}\epsilon^{\nu]\sigma\tau\gamma}u_{\gamma}u^{\rho}$ & $w_{18}$ & $u^{\mu}\Delta^{\rho[\sigma}\epsilon^{\tau]\lambda\nu\gamma}u_{\gamma}$\tabularnewline
\hline
$w_{19}$ & \multicolumn{5}{c|}{$\Delta^{\mu\rho}u^{[\tau}\Delta^{\sigma][\lambda}u^{\nu]}-\Delta^{\mu[\tau}u^{\sigma]}u^{[\lambda}\Delta^{\nu]\rho}$}\tabularnewline
\hline
$w_{20}$ & \multicolumn{5}{c|}{$\Delta^{\mu[\nu}\epsilon^{\lambda]\rho\gamma[\tau}u^{\sigma]}u_{\gamma}+(\Delta^{\rho[\lambda}u^{\nu]}\epsilon^{\mu\sigma\tau\gamma}+\Delta^{\mu\rho}u^{[\lambda}\epsilon^{\nu]\sigma\tau\gamma})u_{\gamma}/2$}\tabularnewline
\hline
$w_{21}$ & \multicolumn{5}{c|}{$\Delta^{\mu[\nu}\epsilon^{\lambda]\rho\gamma[\tau}u^{\sigma]}u_{\gamma}-(\Delta^{\rho[\lambda}u^{\nu]}\epsilon^{\mu\sigma\tau\gamma}+\Delta^{\mu\rho}u^{[\lambda}\epsilon^{\nu]\sigma\tau\gamma})u_{\gamma}/2$}\tabularnewline
\hline
$w_{22}$ & \multicolumn{5}{c|}{$\Delta^{\mu\rho}\epsilon^{\nu\lambda\gamma[\sigma}u^{\tau]}u_{\gamma}-\Delta^{\mu[\nu}\epsilon^{\lambda]\rho\gamma[\tau}u^{\sigma]}u_{\gamma}-\epsilon^{\mu\nu\lambda\gamma}u_{\gamma}u^{[\sigma}\Delta^{\tau]\rho}/3$}\tabularnewline
\hline
$w_{23}$ & \multicolumn{5}{c|}{$(\Delta^{\rho[\lambda}u^{\nu]}\epsilon^{\mu\sigma\tau\gamma}-\Delta^{\mu\rho}u^{[\lambda}\epsilon^{\nu]\sigma\tau\gamma})u_{\gamma}/2-\Delta^{\mu[\nu}u^{\lambda]}\epsilon^{\sigma\tau\gamma\rho}u_{\gamma}/3$}\tabularnewline
\hline
$w_{24}$ & \multicolumn{5}{c|}{$\Delta^{\mu\rho}\Delta^{\lambda[\sigma}\Delta^{\tau]\nu}+\epsilon^{\mu\lambda\nu\gamma}u_{\gamma}
\epsilon^{\rho\sigma\tau\kappa}u_{\kappa}/6-\Delta^{\mu[\nu}\Delta^{\lambda][\sigma}\Delta^{\tau]\rho}$}
\tabularnewline
\hline
\end{tabular}

\caption{Irreducible decompositions of $W^{\mu\lambda\nu\rho\sigma\tau}$.}\label{Table-W}
\end{table}
%

\section{Matching conditions and semi-positivity constraints}\label{sec:matching-condition}

Tables~\ref{Table-G-I-O-F}--\ref{Table-W} introduce 98 undetermined factors, leading to very complicated 
forms of the dissipative currents. Fortunately, some of these factors can be eliminated by imposing matching 
conditions and the semi-positivity of the entropy production rate $\partial_\mu s^\mu\geq 0$. 
In this work, we impose the following matching conditions:
\begin{align}\label{Matching-conditions}
u_{\mu}(\delta T^{\mu\nu}_{S}+\delta T^{\mu\nu}_{A})=0,~~~u_{\mu}\delta j^{\mu}=0,~~~u_{\mu}\delta S^{\mu\lambda\nu}=0.
\end{align}
By substituting Eqs.~(\ref{EMTsymmetricgradientdecomposition})-(\ref{Spingradientdecomposition}) into 
(\ref{Matching-conditions}) and applying the irreducible decompositions in Tables~\ref{Table-G-I-O-F}-\ref{Table-W}, 
the conditions are converted to, 
\begin{align}
u_{\mu}\left[\delta T^{\mu\nu}_{S}+\delta T^{\mu\nu}_{A}\right]=0&\quad\implies\quad
\begin{cases}
     &h_{1}+n_{1}=0,\quad h_{5}=h_{6}=0,\quad k_{1}-p_1=0,\quad k_{3}=0,\\
     &l_{1}+q_{1}=0,\quad l_{2}+{q_{4}}=0,\quad m_{1}+r_{1}=0,\quad m_{2}+r_2=0,\\
     &m_5+r_3=0,\quad m_6-r_4=0,\quad m_7=m_8=0.
\end{cases}\label{matchingconditions-1}
\\
u_{\mu}\delta j^{\mu}=0&\quad \implies\quad 
\hspace{0.65cm} g_2=g_3=0,\quad i_2=0,\quad f_5=f_6=0.\label{matchingconditions-2}
\\
u_{\mu}\delta S^{\mu\lambda\nu}=0&\quad\implies\quad
\begin{cases}
  &t_9=t_{10}=0,\quad u_{5}=u_{6}=0,\quad v_{5}=v_{6}=0,\quad v_{7}=v_{8}=0,\\
  &w_{11}=w_{12}=0,\quad w_{13}=w_{14}=0,\quad w_{15}=w_{16}=0,\quad w_{17}=w_{18}=0.
\end{cases}\label{matchingconditions-3}
\end{align}
After taking these 34 constraints into account, we are left with $98-34=64$ undetermined coefficients.

On the other hand, the second law of thermodynamics requires the semi-positivity of entropy production rate, given in Eq.~\eqref{entropyrate}. In order to simplify this condition, we decompose the thermo-hydrodynamic gradient terms using the irreducible representations of the rotation group, identical to the procedure discussed in the preceding section that led to Tables~\ref{Table-G-I-O-F}-\ref{Table-W}. We start from the symmetric shear tensor,  
\begin{align}\label{thermalsheardecomposition}
    \xi^{\mu\nu}=(D\beta)u^{\mu}u^{\nu}+\frac{1}{3T}\theta\Delta^{\mu\nu}+\frac{1}{T}\mathcal{J}_h^{\alpha}\left(\Delta^\mu_\alpha u^{\nu}+\Delta^\nu_\alpha u^\mu \right)+\frac{1}{T}\sigma^{\alpha\beta} \Delta^{\mu\nu}_{\alpha\beta},
\end{align}
where we have introduced the comoving and spatial derivatives $D\equiv u_\mu\partial^\mu$ and $\nabla^\mu\equiv\Delta^{\mu\nu}\partial_\nu$, the expansion scalar $\theta\equiv \partial_\mu u^\mu$, the heat flow $\mathcal{J}_h^\mu\equiv D u^\mu-(1/T)\nabla^\mu T$, and the shear tensor $\sigma^{\mu\nu}\equiv \Delta^{\mu\nu}_{\alpha\beta}\partial^\alpha u^\beta$.
The following gradient term, namely the difference between the reduced spin potential and the thermal vorticity, $\Omega_{\mu\nu}-\varpi_{\mu\nu}$, is antisymmetric and can be decomposed as,
\begin{align}\label{Omega-omegadecomposition}
\Omega^{\mu\nu}-\varpi^{\mu\nu}=\frac{1}{T}\mathcal{E}^{\alpha}\left(\Delta^\nu_\alpha u^{\mu}-\Delta^\mu_\alpha u^\nu \right)-\frac{1}{2T}\epsilon^{\mu\nu\alpha\beta}u_{\alpha}\mathcal{B}_{\beta},
\end{align}
where the electric-like part is $\mathcal{E}^{\mu}\equiv T u_{\nu}(\Omega^{\nu\mu}-\varpi^{\nu\mu})$, and the magnetic-like part reads $\mathcal{B}^{\mu}\equiv T\epsilon^{\alpha\beta\gamma\mu}u_{\gamma}(\Omega_{\alpha\beta}-\varpi_{\alpha\beta})$. The decomposition of the next gradient contribution, i.e., the derivative of the reduced chemical potential over temperature, is given by 
\begin{align}\label{dzetadecomposition}
    \partial_{\mu}\zeta=\left(D\zeta\right)u_{\mu}+\left(\nabla_{\alpha}\zeta\right) \Delta^\alpha_\mu \;.
\end{align}
Finally, the derivative of the reduced spin potential is decomposed as follows, with details of the derivation relegated to Appendix~\ref{AppendixA},
\begin{align}\label{spinpotentialgradientdecomposition}
 \partial_{\mu}\Omega_{\lambda\nu}=&\ 2\mathcal{X}^{\gamma}u_{\mu}u_{[\lambda}\Delta_{\nu]\gamma}-
\frac12\mathcal{Y}^{\gamma}u_{\mu}\epsilon_{\lambda\nu\sigma\gamma}u^{\sigma}
+\frac12\mathcal{Z}u_{[\lambda}\Delta_{\nu]\mu}+\mathcal{T}^{\gamma}u_{[\lambda}\epsilon_{\nu]\mu\alpha\gamma}u^{\alpha}+2\mathcal{F}^{\rho\sigma}u_{[\lambda}\Delta_{\nu]\sigma}\Delta_{\mu \rho}\nonumber\\
 &\ -\frac18\mathcal{H}\epsilon_{\lambda\nu\alpha\mu}u^{\alpha}-\frac12\mathcal{G}_{[\lambda}\Delta_{\nu]\mu}-\frac12\mathcal{I}^{\rho\sigma}\Delta_{\mu \rho}\epsilon_{\lambda\nu\tau \sigma}u^{\tau}\;.
\end{align}
The scalars, vectors, and tensors appearing in the expression above are defined as follows,
\begin{align}
&\mathcal{X}^{\gamma}=u^{[\rho}\Delta^{\sigma]\gamma}D\Omega_{\rho\sigma},\quad \mathcal{Y}^{\gamma}=\epsilon^{\rho\sigma\tau\gamma}u_{\tau}D\Omega_{\rho\sigma},\quad
\mathcal{Z}=u^{[\rho}\nabla^{\sigma]}\Omega_{\rho\sigma},\quad
\mathcal{T}^{\gamma}=u^{[\rho}\epsilon^{\sigma]\lambda\theta\gamma}u_{\lambda}\partial_{\theta}\Omega_{\rho\sigma},     \nonumber\\
&\mathcal{F}^{\rho\sigma}=u^{[\gamma}\Delta^{\theta](\rho}\nabla^{\sigma)}\Omega_{\gamma\theta}-\Delta^{\rho\sigma}\mathcal{Z}/4
,\quad
\mathcal{H}=\epsilon^{\rho\sigma\lambda\gamma}u_{\lambda}\partial_{\gamma}\Omega_{\rho\sigma},\quad
\mathcal{G}^{\gamma}=2\Delta^{\gamma\tau}\nabla^{\rho}\Omega_{\rho\tau},    \nonumber\\
&\mathcal{I}^{\rho\sigma}=u_{\gamma}\epsilon^{\theta\lambda\gamma(\rho}\nabla^{\sigma)}\Omega_{\theta\lambda}-\Delta^{\rho\sigma}\mathcal{H}/4.
\end{align}
Note that the tensors $\mathcal{F}^{\mu\nu}$ and $\mathcal{I}^{\mu\nu}$ are symmetric, but not traceless, with their traces being related to $\mathcal{Z}$ and $\mathcal{H}$, respectively. They can be further written in terms of traceless symmetric tensors $\mathcal{F}^{\mu\nu}_{S}$ and $\mathcal{I}^{\mu\nu}_{S}$ as
\begin{align}
    \mathcal{F}^{\mu\nu}=\mathcal{F}^{\mu\nu}_{S}+\frac{1}{12}\Delta^{\mu\nu}\mathcal{Z},\quad
    \mathcal{I}^{\mu\nu}=\mathcal{I}^{\mu\nu}_{S}+\frac{1}{12}\Delta^{\mu\nu}\mathcal{H}.
\end{align}
Then by contracting the above gradient terms~\eqref{thermalsheardecomposition}-\eqref{spinpotentialgradientdecomposition} with the dissipative currents~\eqref{EMTsymmetricgradientdecomposition}--\eqref{Spingradientdecomposition}, and applying the matching conditions \eqref{matchingconditions-1}-\eqref{matchingconditions-3}, we arrive at the following results
\begin{align}\label{contraction2}
&T^2\xi_{\mu\nu}\delta T^{\mu\nu}_{S}=\,h_1 \mathcal{J}_h^\mu \mathcal{J}_{h,\mu}+h_2\theta^2+h_3\sigma^{\mu\nu}\sigma_{\mu\nu}+\left(h_4 T D\beta+k_2D\zeta+m_3\mathcal{Z}+m_4\mathcal{H}\right)\theta
\nonumber\\
&\hspace{2.1cm}-\mathcal{J}^h_\mu\left(- k_1\nabla^\mu \zeta+l_1 \mathcal{E}^\mu  +l_2 \mathcal{B}^\mu +m_1\mathcal{X}^\mu  +m_2\mathcal{Y}^\mu +m_5\mathcal{G}^\mu +m_6\mathcal{T}^\mu \right)
\nonumber\\
&\hspace{2.1cm}+\sigma_{\mu\nu}\left(m_9\mathcal{I}_S^{\mu\nu}+m_{10}\mathcal{F}_S^{\mu\nu}\right) \,,\\ 
&T^2\partial_{\mu}\zeta\delta j^{\mu}=\,i_1(\nabla^\mu\zeta)(\nabla_\mu\zeta)-\left(-g_1\mathcal{J}_h^{\mu}+o_1\mathcal{E}^{\mu}+o_2\mathcal{B}^{\mu}+f_1\mathcal{X}^{\mu}+f_2\mathcal{Y}^{\mu}+f_3\mathcal{G}^{\mu}+f_4\mathcal{T}^{\mu}\right)\nabla_{\mu}\zeta\,,\\
&T^2(\Omega_{\mu\nu}-\varpi_{\mu\nu})\delta T^{\mu\nu}_{A}=\,-q_{1}\mathcal{E}^\mu \mathcal{E}_\mu-q_{2}\mathcal{B}^\mu \mathcal{B}_\mu-(q_3+q_4)\mathcal{E}^{\mu}\mathcal{B}_{\mu}\nonumber\\
&\hspace{3.6cm}-\mathcal{E}_{\mu}\left(-n_1\mathcal{J}_h^{\mu}+p_1\nabla^{\mu}\zeta+r_1\mathcal{X}^{\mu}+r_2\mathcal{Y}^{\mu}+r_3\mathcal{G}^{\mu}-r_4\mathcal{T}^{\mu}\right)
\nonumber\\
&\hspace{3.6cm}-\mathcal{B}_{\mu}\left(-n_{2}\mathcal{J}_h^{\mu}+p_2\nabla^{\mu}\zeta+r_5\mathcal{X}^{\mu}+r_6\mathcal{Y}^{\mu}+r_7\mathcal{G}^{\mu}-r_8\mathcal{T}^{\mu}\right)\,,\\
&T^2\partial_{\mu}\Omega_{\lambda\nu}\delta S^{\mu\lambda\nu}=\,(t_1 TD\beta+t_2\theta+u_1 D\zeta)\mathcal{Z}+(t_4 TD\beta+t_5\theta+u_3 D\zeta)\mathcal{H}
+w_1\mathcal{Z}^{2}+w_2\mathcal{H}^{2}+2w_3\mathcal{Z}\mathcal{H}
\nonumber\\
&\hspace{2.7cm}-(t_3\mathcal{J}^h_{\mu}+u_2\nabla_{\mu}\zeta+v_2\mathcal{B}_{\mu}+v_3\mathcal{E}_{\mu}+w_7\mathcal{X_{\mu}}+w_8\mathcal{Y}_{\mu})\mathcal{G}^{\mu}
\nonumber\\
&\hspace{2.7cm}-(t_6\mathcal{J}^h_{\mu}+u_4\nabla_{\mu}\zeta+v_1\mathcal{B}_{\mu}+v_4\mathcal{E}_{\mu}+w_5\mathcal{X_{\mu}}+w_6\mathcal{Y}_{\mu})\mathcal{T}^{\mu}
\nonumber\\
&\hspace{2.7cm}-w_9\mathcal{G}^{\mu}\mathcal{G}_{\mu}-w_{19}\mathcal{T}^{\mu}\mathcal{T}_{\mu}-w_{21}\mathcal{G}^{\mu}\mathcal{T}_{\mu}\nonumber\\
&\hspace{2.7cm}+(t_7\mathcal{F}^{\mu\nu}_{S}+t_8I^{\mu\nu}_{S})\sigma_{\mu\nu}+w_{10}\mathcal{F}^{\mu\nu}_{S}\mathcal{F}_{S,\mu\nu}+(w_{22}+w_{23})\mathcal{I}^{\mu\nu}_{S}\mathcal{F}_{S,\mu\nu}-\frac{1}{2}w_{24}\mathcal{I}^{\mu\nu}_{S}\mathcal{I}_{S,\mu\nu}\,.
\label{contraction4}
\end{align}
The total entropy production rate, $\partial_{\mu}s^{\mu}$, is then obtained by substituting these results into Eq.~\eqref{entropyrate}. 

\begin{table}[H]
\renewcommand{\arraystretch}{1.5}
\begin{tabularx}{1\textwidth} { 
| >{\raggedright\arraybackslash}X 
| >{\raggedright\arraybackslash}X
| >{\raggedright\arraybackslash}X
| >{\raggedright\arraybackslash}X
| >{\centering\arraybackslash}X 
| >{\raggedleft\arraybackslash}X | }
\hline
Scalars & Pseudo-scalars & Vectors & Pseudo-vectors & Tensors & Pseudo-tensors \\
 
\hline
$\theta$  & $\mathcal{H}$  &  $\mathcal{J}_h^\mu$ & $\mathcal{B}^{\mu}$ & $\sigma^{\mu\nu}$ & $\mathcal{I}^{\mu\nu}_{S}$ \\
\hline
$TD\beta$ & ---- & $\nabla^{\mu}\zeta$  & $\mathcal{Y}^{\mu}$ & $\mathcal{F}^{\mu\nu}_{S}$ & --- \\
\hline
$ D\zeta$ & ---- &$ \mathcal{X}^{\mu} $& $\mathcal{T}^{\mu}$& --- &----\\
\hline
$\mathcal{Z}$ & ---- & $\mathcal{E}^{\mu}$ & --- & ---& ---\\
\hline
--- & ----- &  $\mathcal{G}^{\mu}$ & --- & --- & ---\\
\hline
\end{tabularx}
\caption{Classification of the thermo-hydrodynamic gradient terms based on their properties under reflections and rotations.}
\label{Table-classification}
\end{table}

We note that $\partial_{\mu}s^{\mu}$ is a Lorentz scalar which is invariant under reflections as well as spatial rotations in the rest frame associated to $u^\mu$. Regarding the thermo-hydrodynamic gradient terms appearing in Eqs. \eqref{contraction2}-\eqref{contraction4}, we may classify them according to their behavior under this class of transformations, as listed in Table~\ref{Table-classification}. In order to ensure all of the terms appearing in the entropy production rate are (parity-even) scalars, some of the coefficients in Eqs. \eqref{contraction2}-\eqref{contraction4} have to be parity-odd while others are parity-even. For simplicity, we set all parity-odd coefficients to zero. This further eliminates 28 of the coefficients, bringing down the total number to $64-28=36$. Then, we can use the information in Table~\ref{Table-classification} to divide the entropy production rate $\partial_{\mu}s^{\mu}$ into the following six terms,
\begin{align}
    \partial_{\mu}s^{\mu}= \partial_{\mu}s^{\mu}_{S.S}+ \partial_{\mu}s^{\mu}_{PS.PS}+\partial_{\mu}s^{\mu}_{V.V}+ \partial_{\mu}s^{\mu}_{PV.PV}+\partial_{\mu}s^{\mu}_{T.T}+ \partial_{\mu}s^{\mu}_{PT.PT}\;.
\end{align}
Here, $\partial_{\mu}s^{\mu}_{S.S}$ and $\partial_{\mu}s^{\mu}_{PS.PS}$ denote the contributions bilinear in scalar and pseudo-scalar fields, respectively, and other terms are labeled in a similar way, with the correspondences 
$S$: scalar, $PS$: pseudo-scalar, $V$: vector, $PV$: pseudo-vector, $T$: tensor, and $PT$: pseudo-tensor. Explicitly, one can write each of these contributions in matrix form as

\begin{align}
\partial_\mu s_{S.S}^\mu&=\frac{1}{T^2}
\begin{pmatrix} 
\theta \\ TD\beta \\ D\zeta \\ \mathcal{Z}
\end{pmatrix}^\text{T}
\left(
\begin{array}{cccc}
 h_2 & h_4/2 & k_2/2 &(2m_3-t_2)/4 \\
 h_4/2 & 0 & 0 & -t_1/4 \\
 k_2/2 & 0 & 0 & -u_1/4\\
 (2m_3-t_2)/4 & -t_1/4& -u_1/4 & -w_1/2 \\
\end{array}
\right)
\begin{pmatrix} 
\theta \\ TD\beta \\ D\zeta \\ \mathcal{Z}
\end{pmatrix}\;,
\\
\partial_\mu s_{PS.PS}^\mu&=-\frac{1}{2T^2}w_2 \mathcal{H}^2\;,
\\
\partial_\mu s_{V.V}^\mu&=-\frac{1}{T^2}
\begin{pmatrix} 
\mathcal{J}_h^{\mu}\\ \nabla^{\mu}\zeta\\ \mathcal{E}^{\mu}\\ \mathcal{G}^{\mu}\\ \mathcal{X}^{\mu}
\end{pmatrix}^\text{T}
\left(
\begin{array}{ccccc}
 -h_{1} &(g_1-k_1)/2 & (l_1-n_1)/2& (2m_5-t_3)/4 & m_{1}/2 \\
 (g_1-k_1)/2& i_1 & (p_1-o_1)/2  &-(2f_3+u_2)/4& -f_1/2 \\
 (l_1-n_1)/2 & (p_1-o_1)/2 & q_1& (2r_3-v_3)/4 & r_1/2  \\
 (2m_5-t_3)/4 & -(2f_3+u_2)/4 & (2r_3-v_3)/4 & -w_9/2 & -w_7/4 \\
  m_1/2 & -f_1/2 & r_1/2 & -w_7/4  & 0\\
\end{array}
\right)
\begin{pmatrix} 
\mathcal{J}^h_{\mu}\\ \nabla_{\mu}\zeta\\ \mathcal{E}_{\mu}\\ \mathcal{G}_{\mu}\\ \mathcal{X}_{\mu}
\end{pmatrix}\;,
\\
\partial_\mu s_{PV.PV}^\mu&=-\frac{1}{T^2}\begin{pmatrix}
\mathcal{B}^{\mu}\\ \mathcal{T}^{\mu}\\ \mathcal{Y}^{\mu}
\end{pmatrix}^\text{T}
\left(
\begin{array}{ccc}
 q_2  & -(2r_8+v_1)/4& r_6/2 \\
 -(2r_8+v_1)/4& -w_{19}/2 & -w_6/4  \\
  r_6/2 & -w_6/4 & 0 \\
\end{array}
\right)
\begin{pmatrix}
\mathcal{B}^{\mu}\\ \mathcal{T}^{\mu}\\ \mathcal{Y}^{\mu}
\end{pmatrix}\;,
\\
\partial_\mu s_{T.T}^\mu&=\frac{1}{T^2}\begin{pmatrix}
\sigma^{\mu\nu}\\ \mathcal{F}^{\mu\nu}_{S}
\end{pmatrix}^\text{T}
\left(
\begin{array}{ccc}
 h_3 & (2m_{10}-t_7)/4 \\
 (2m_{10}-t_7)/4& -w_{10}/2\\
\end{array}
\right)
\begin{pmatrix}
\sigma^{\mu\nu}\\ \mathcal{F}^{\mu\nu}_{S}
\end{pmatrix}\;,
\\
\partial_\mu s_{PT.PT}^\mu&=\frac{1}{4T^2}w_{24}\mathcal{I}^{\mu\nu}_{S}\mathcal{I}_{\mu\nu\, S}\;.
\end{align}
The semi-positivity of $\partial_\mu s^\mu$ requires each of these parts to be semi-positive. There are two equivalent approaches to establish semi-positivity: (1) diagonalize the relevant matrices such that the results are expressed in terms of perfect square formulas, or (2) use the fact that all eigenvalues of a matrix are non-negative if and only if all principal minors of the matrix are non-negative. In this work we use the second approach, which leads to the following conditions,
\begin{align}
\label{eq:divS_s}
   \partial_{\mu}s^{\mu}_{S.S}\geq0\quad\implies\quad &\begin{cases}
        &h_2\geq0,\quad w_1\leq 0,\quad h_4=k_2=t_1=u_1=0,\\
        &-8h_2w_1\geq (2m_3-t_2)^2,
    \end{cases}
\\
    \partial_{\mu}s^{\mu}_{PS.PS}\geq0\quad\implies\quad &
       \hspace{0.7cm}w_2\leq 0,
\\
    \partial_{\mu}s^{\mu}_{V.V}\geq0\quad\implies\quad&\begin{cases}
    &{m_1=f_1=r_1=w_7=0},\quad h_1\leq 0,~i_1\geq 0,~w_9\leq 0,\\
       &q_1+h_1=0,\quad g_1-o_1=0,\quad t_3+v_3=0,\\
       &-4h_1 i_1\geq(k_1-g_1)^2,\quad 8h_1w_9\geq(2m_5-t_3)^2,\quad -8i_1w_9\geq (2f_3+u_2)^2,\\
       & h_1(2f_3+u_2)^2-i_1(2m_5-t_3)^2+2w_9(k_1-g_1)^2\\
       & \hspace{0.7cm}+(2f_3+u_2)(2m_5-t_3)(k_1-g_1)+8h_1i_1w_9\geq0,
\end{cases}
\\
    \partial_{\mu}s^{\mu}_{PV.PV}\geq0\quad\implies\quad&\begin{cases}
        &{q_2\geq 0},\quad {w_{19}\leq 0},\quad {r_6=w_6=0},\\
        &-8q_2w_{19}\geq (2r_8+v_1)^2,
    \end{cases}
\\
    \partial_{\mu}s^{\mu}_{T.T}\geq0\quad\implies\quad&\begin{cases}
        &{h_{3}\geq 0},\quad {w_{10}\leq 0},\\
        &-8h_3w_{10}\geq (2m_{10}-t_7)^2,
    \end{cases}
\\
\label{eq:divS_PT}
    \partial_{\mu}s^{\mu}_{PT.PT}\geq0\quad\implies\quad&\hspace{0.7cm}{w_{24}\geq 0}.
\end{align}
These constraints include 13 equalities, which reduce the number of independent coefficients to $36-13=23$. The remaining quantities are further constrained by the inequalities introduced in Eqs. \eqref{eq:divS_s}--\eqref{eq:divS_PT}, which can be classified as follows: 11 inequalities demand that a certain coefficient has a definite sign, 6 inequalities set a bound for the product of two coefficients, and 1 inequality constrains the product of three coefficients. Taking all of these into account, we can construct the final form of the dissipative currents, as will be done in the following.

\section{Final expressions and comparison with previous results}\label{sec:final-result}

In this section we provide the explicit expressions for the dissipative currents at first 
order in gradients. The symmetric dissipative part of the energy-momentum tensor takes the 
following form:
\begin{align}
\delta T^{\mu\nu}_{S}=&2u^{(\mu}\left[-\kappa_{h}(\mathcal{J}_h^{\nu)}-\mathcal{E}^{\nu)})+\kappa_{\zeta}\,\nabla^{\nu)}\zeta+\kappa_{\mathcal{G}}\,\mathcal{G}^{\nu)}\right]
+\left(\zeta_b\theta+\zeta_{\mathcal{Z}}\,\mathcal{Z}\right)\Delta^{\mu\nu}+2\left(\eta\sigma^{\mu\nu}
+\eta_{\mathcal{F}}\,\mathcal{F}_{S}^{\mu\nu}\right),
\label{EMTsymmetricdissipativecurrent}
\end{align}
where the thermo-hydrodynamic fields, the corresponding transport coefficients, their relations with 
factors introduced in Tables~\ref{Table-G-I-O-F}-\ref{Table-W}, and the constraints and physical interpretations of these coefficients 
are given in Table~\ref{Tablesymmetric}.

\begin{table}[H]\label{symmset}
\renewcommand{\arraystretch}{2.3}
\begin{tabularx}{1\textwidth} { 
| >{\raggedright\arraybackslash}X
| >{\raggedright\arraybackslash}X
| >{\raggedright\arraybackslash}X
| >{\centering\arraybackslash}X 
| >{\raggedleft\arraybackslash}X | }
\hline
\textbf{Thermo-hydrodynamic field} &  \textbf{Coefficient} & \textbf{Constraint} & \textbf{Interpretation}\\
\hline
$\theta=\nabla^{\mu}u_{\mu}$&$\zeta_b={h_2}/{T}$&$\zeta_b \geq 0$& Bulk viscosity\\
\hline
$\sigma^{\mu\nu}=\nabla^{(\mu}u^{\nu)}-\frac{1}{3}\theta\Delta^{\mu\nu}$&$\eta={h_{3}}/{(2T)}$&$\eta\geq 0$&Shear viscosity\\
\hline
$\mathcal{Z}=u^{[\mu}\nabla^{\nu]}\Omega_{\mu\nu}$  &  $\zeta_{\mathcal{Z}}=m_3/T$ &$4\zeta\chi_\mathcal{Z}\geq(\zeta_\mathcal{Z}+\chi_{\theta})^{2}$&{Gyro-bulk viscosity}\\
\hline
$\mathcal{F}^{\mu\nu}_{S}={-u_\rho \nabla^{(\mu}\Omega^{\nu)\rho}-\frac{1}{3}\Delta^{\mu\nu}\mathcal{Z}}$ & $\eta_{\mathcal{F}}={m_{10}}/{(2T)}$ &$16\eta\chi_\mathcal{F}\geq(4\eta_\mathcal{F}+\chi_{\sigma})^{2}$&Gyro-shear viscosity\\
\hline
\end{tabularx}
\caption{Summary of contributions to \( \delta T^{\mu\nu}_{S} \) in Eq.~\eqref{EMTsymmetricdissipativecurrent} 
from various thermo-hydrodynamic fields, along with the corresponding transport coefficients. The 
first three coefficients in Eq.~\eqref{EMTsymmetricdissipativecurrent} are not independent, as they also appear in 
\(\delta T^{\mu\nu}_{A}\) in Eq.~\eqref{EMTantisymmetricdissipativecurrent} due to the matching 
condition~\eqref{Matching-conditions}; they are listed in Table \ref{Tableantisymmetric}.}
\label{Tablesymmetric}
\end{table}

In Table \ref{Tablesymmetric}, $\eta$ and $\zeta_b$ are the shear and bulk viscosities. Two new 
transport coefficients, $\zeta_{\mathcal{Z}}$ and $\eta_{\mathcal{F}}$, are identified for 
the symmetric part of the stress-energy tensor, corresponding to the gradients of the spin potential.
The first three coefficients in Eq. \eqref{EMTsymmetricdissipativecurrent} multiplying
vector terms also appear in $\delta T^{\mu\nu}_{A}$, according to
Eq.~\eqref{EMTantisymmetricdissipativecurrent}, because they are involved in the matching 
condition $u_\mu (T^{\mu\nu}_S+T^{\mu\nu}_A)=0$ in Eq.~\eqref{Matching-conditions}; they have
been omitted in Table \ref{Tablesymmetric} and quoted in Table \ref{Tableantisymmetric}.

The anti-symmetric contribution to the energy-momentum tensor reads:
\begin{align}
\delta T^{\mu\nu}_{A}=&-2u^{[\mu}\left[-\kappa_{h}(\mathcal{J}_h^{\nu]}-\mathcal{E}^{\nu]})+\kappa_{\zeta}\,\nabla^{\nu]}\zeta+\kappa_{\mathcal{G}}\,\mathcal{G}^{\nu]}\right]-2\gamma_{\phi}\phi^{\mu\nu}-2\gamma_{\Xi}\Xi^{\mu\nu}\;,
\label{EMTantisymmetricdissipativecurrent}
\end{align}
and the relevant transport coefficients are summarized in Table~\ref{Tableantisymmetric}. 
The coefficients $\kappa_{h}$ and $\gamma_{\phi}$ have been previously obtained 
in Ref.~\cite{Hattori:2019lfp} and named boost heat conductivity 
and rotational viscosity, denoted as $\lambda$ and $\gamma$ 
respectively. It should be noted that the current associated with the coefficient 
$\kappa_{h}$ does not precisely match that of $\lambda$ in Ref. \cite{Hattori:2019lfp}
due to the choice of the matching condition. Nevertheless, they basically have the same physical 
meaning. Studies on the rotational viscosity coefficient have been also conducted in Refs.~\cite{Hongo:2022izs,Hidaka:2023oze}. We thus identify three new transport coefficients: $\kappa_{\zeta}$, $\kappa_{\mathcal{G}},$ and $\gamma_{\Xi}$, which appear due to the inclusion of the gradient of the spin 
potential and the finite chemical potential. 

\begin{table}[H]
\renewcommand{\arraystretch}{2.1}
\begin{tabularx}{1\textwidth} { 
| >{\raggedright\arraybackslash}X 
| >{\raggedright\arraybackslash}X
| >{\raggedright\arraybackslash}X
| >{\centering\arraybackslash}X 
| >{\raggedleft\arraybackslash}X | }
\hline
\textbf{Thermo-hydrodynamic field} &  \textbf{Coefficient} & \textbf{Constraint} & \textbf{Interpretation}\\
\hline
$\mathcal{J}_h^{\mu}-\mathcal{E}^{\mu}= Du^{\mu}-\frac{1}{T}\nabla^{\mu}T+T(\Omega^{\mu\nu}-\varpi^{\mu\nu})u_{\nu}$&$\kappa_{h}=-h_{1}/(2T)$&$\kappa_{h}\geq 0$& Boost heat conductivity\\
\hline
$\nabla^{\mu}\zeta$ &  $\kappa_{\zeta}={k_1}/{(2T)}$ &$8T\kappa_h\mathcal{K}_\zeta\geq(T\mathcal{K}_h-2\kappa_\zeta)^2$& Chemical-boost heat conductivity\\
\hline
$\mathcal{G}^{\mu}=2\Delta^{\mu\tau}\nabla^{\rho}\Omega_{\rho\tau}$&$\kappa_{\mathcal{G}}=-m_5/(2T)$ &$16\kappa_h\chi_\mathcal{G}\geq(8\kappa_\mathcal{G}-\chi_h)^2$& Gyro-boost heat conductivity\\
\hline
$\phi^{\mu\nu}=T\Delta^{\mu[\sigma}\Delta^{\rho]\nu}(\Omega_{\rho\sigma}-\varpi_{\rho\sigma})$ &  $\gamma_{\phi}=q_2/T$ &$\gamma_{\phi}\geq0$& Rotational viscosity\\
\hline
$\Xi^{\mu\nu}=-u_\rho\nabla^{[\mu}\Omega^{\nu]\rho}$&$\gamma_{\Xi}={r_{8}}/{T}$&
$16\gamma_\phi\chi_\Xi\geq (4\gamma_\Xi+\chi_\phi)^2$&
Gyro-rotation viscosity\\
\hline
\end{tabularx}
\caption{Summary of contributions to $\delta T^{\mu\nu}_{A}$ in Eq.~\eqref{EMTantisymmetricdissipativecurrent} 
from various thermo-hydrodynamic fields, along with the corresponding transport coefficients.}
\label{Tableantisymmetric}
\end{table}

%
%
For the conserved vector current, we obtain the following expression:
\begin{align}
\delta j^{\mu}=\mathcal{K}_{\zeta}\nabla^{\mu}\zeta+\mathcal{K}_{h}\left(\mathcal{J}_h^{\mu}-\mathcal{E}^{\mu}\right)+\mathcal{K}_{\mathcal{G}}\,\mathcal{G}^{\mu}.
\label{particlenodissipativecurrent}
\end{align}
A summary of the corresponding transport coefficients, along with their respective constraints and physical interpretations, is provided in Table~\ref{taparticlenumber}. The coefficient $\mathcal{K}_\zeta$ relates the dissipative current to the gradient of the chemical
potential and can then be identified as the diffusion coefficient. However, in the context of relativistic spin hydrodynamics, we observe the appearance of two additional transport coefficients, $\mathcal{K}_{h}$ and $\mathcal{K}_{\mathcal{G}}$, which are not present in the traditional formulation without the spin tensor. 

\begin{table}[H]
\renewcommand{\arraystretch}{2.9}
\begin{tabularx}{1\textwidth} { 
| >{\raggedright\arraybackslash}X 
| >{\raggedright\arraybackslash}X
| >{\raggedright\arraybackslash}X
| >{\centering\arraybackslash}X 
| >{\raggedleft\arraybackslash}X | }
\hline
\textbf{Thermo-hydrodynamic field} &  \textbf{Coefficient} & \textbf{Constraint} & \textbf{Interpretation}\\
\hline
$\nabla^{\mu}\zeta$ & $\mathcal{K}_{\zeta}={i_1}/{T^{2}}$& $\mathcal{K}_{\zeta}\geq0$& Diffusion coefficient\\
 \hline
$\mathcal{J}_h^{\mu}-\mathcal{E}^{\mu}=Du^{\mu}-\frac{1}{T}\nabla^{\mu}T+T(\Omega^{\mu\nu}-\varpi^{\mu\nu})u_{\nu}$ &$\mathcal{K}_{h}={g_1}/{T^2}$&$8T\kappa_h\mathcal{K}_\zeta\geq(T\mathcal{K}_h-2\kappa_\zeta)^2$ &Diffusion-boost heat conductivity\\
\hline
$\mathcal{G}^{\mu}=2\Delta^{\mu\tau}\nabla^{\rho}\Omega_{\rho\tau}$ & $\mathcal{K}_{\mathcal{G}}=-{f_3}/{T^{2}}$&$8T\chi_G\mathcal{K}_\zeta\geq(2T\mathcal{K}_\mathcal{G}+\chi_\zeta)^2$&Diffusion-gyro-boost heat conductivity\\
\hline
\end{tabularx}
\caption{Summary of various transport coefficients for \( \delta j^{\mu} \) in Eq.~\eqref{particlenodissipativecurrent}.}
\label{taparticlenumber}
\end{table}

Finally, we report the expression for the spin dissipative current: 
\begin{align}
\delta S^{\mu\lambda\nu}=&\frac{1}{T}\left[\Delta^{\mu[\lambda}\left(\chi_{h}(\mathcal{J}_h^{\nu]}-\mathcal{E}^{\nu]})+\chi_{\zeta}\nabla^{\nu]}\zeta+\chi_{\mathcal{G}}\mathcal{G}^{\nu]}\right)+\Delta^{\mu[\lambda}u^{\nu]}\left(\chi_{\theta}\theta+\chi_{\mathcal{Z}}\mathcal{Z}\right)+\left(\chi_{\sigma}\sigma^{\mu[\lambda}u^{\nu]}+\chi_{\mathcal{F}}\mathcal{F}_{S}^{\mu[\lambda}u^{\nu]}\right)\right.\nonumber\\
&\left.+\left(\chi_{\phi}\phi^{\mu[\lambda}u^{\nu]}+\chi_{\Xi}\Xi^{\mu[\lambda}u^{\nu]}\right)+\left(\chi_{\mathcal{H}}\epsilon^{\mu \lambda\nu\rho}u_{\rho}\mathcal{H}+\chi_{\mathcal{I}}\mathcal{I}^{\mu}_{\rho,S}\epsilon^{\lambda\nu\rho\gamma}u_{\gamma}\right)\right]\;,
\label{Spindissipativecurrent}
\end{align}
with transport coefficients listed in Table~\ref{Tablespin}. The coefficients \( \chi_{Z}, \chi_{F}, \chi_{\Xi} \) were identified in Refs.~\cite{She:2021lhe,Biswas:2023qsw,Drogosz:2024gzv}; most coefficients related to the derivatives
of the spin potential were obtained in Refs.~\cite{Gallegos:2020otk,Dey:2024cwo}. In addition to the 
constraints in Tables~\ref{Tablesymmetric}-\ref{Tablespin}, we have another requirement:
\begin{equation}
16T\kappa_h \mathcal{K}_\zeta \chi_\mathcal{G}\geq T\mathcal{K}_\zeta(8\kappa_\mathcal{G}-\chi_h)^2+2\kappa_h(2T\mathcal{K}_\mathcal{G}+\chi_\zeta)^2+2\chi_\mathcal{G}(2\chi_\zeta-T\mathcal{K}_h)^2-(8\kappa_\mathcal{G}-\chi_h)(2T\mathcal{K}_\mathcal{G}+\chi_\zeta)(2\chi_\zeta-T\mathcal{K}_h)\;.
\end{equation}
\begin{table}[H]
\renewcommand{\arraystretch}{2.5}
\begin{tabularx}{1\textwidth} { 
| >{\raggedright\arraybackslash}X 
| >{\raggedright\arraybackslash}X
| >{\raggedright\arraybackslash}X
| >{\centering\arraybackslash}X 
| >{\raggedleft\arraybackslash}X | }
\hline
\textbf{Thermo-hydrodynamic field}& \textbf{Coefficient} & \textbf{Constraint}& \textbf{Interpretation}\\
 \hline
$(\mathcal{J}_h^{\nu}-\mathcal{E}^{\nu})=D u^{\nu}-\frac{1}{T}\nabla^{\nu}T+T(\Omega^{\nu\lambda}-\varpi^{\nu\lambda})u_{\lambda}$& $\chi_{h}=-t_3/T$& $16\kappa_{h}\chi_\mathcal{G}\geq(8\kappa_\mathcal{G}-\chi_{h})^{2}$& Spin-boost heat conductivity\\
\hline
$\nabla^{\nu}\zeta$&$\chi_{\zeta}=-u_2/T$&$8T\chi_\mathcal{G}\mathcal{K}_{\zeta}\geq(2T\mathcal{K}_\mathcal{G}+\chi_{\zeta})^{2}$& Spin-chemical-boost heat conductivity\\
\hline
$\mathcal{G}^{\nu}=2\Delta^{\nu\tau}\nabla^{\rho}\Omega_{\rho\tau}$& $\chi_{\mathcal{G}}=-w_9/T$&$\chi_{\mathcal{G}}\geq 0$&  Spin-gyro-boost heat conductivity\\
\hline
$\theta=\nabla^{\mu}u_{\mu}$ & $\chi_{\theta}=-t_2/(2T)$&$4\zeta\chi_\mathcal{Z}\geq(\zeta_\mathcal{Z}+\chi_{\theta})^{2}$ &{  Spin-bulk viscosity}\\
\hline
$\mathcal{Z}=u^{[\mu}\nabla^{\nu]}\Omega_{\mu\nu}$ & $\chi_{\mathcal{Z}}=-w_1/(2T)$&$\chi_{\mathcal{Z}}\geq 0$&{ Spin-gyro-bulk viscosity}\\
\hline
$\sigma^{\mu\lambda}=\nabla^{(\mu}u^{\lambda)}-\frac{1}{3}\theta\Delta^{\mu\lambda}$& $\chi_{\sigma}=-t_7/T$&$16\eta\chi_\mathcal{F}\geq(4\eta_\mathcal{F}+\chi_{\sigma})^{2}$&{ Spin-shear viscosity}\\
\hline
$\mathcal{F}^{\mu\lambda}_{S}=-u_{\rho} \nabla^{(\mu}\Omega^{\lambda)\rho}-\frac{1}{3}\Delta^{\mu\lambda}\mathcal{Z}$ & $\chi_{\mathcal{F}}=-w_{10}/T$& $\chi_{\mathcal{F}}\geq 0$&{ Spin-gyro-shear viscosity}\\
\hline
$\phi^{\mu\lambda}=T\Delta^{\mu[\sigma}\Delta^{\rho]\lambda}(\Omega_{\rho\sigma}-\varpi_{\rho\sigma})$& $\chi_{\phi}=2v_1/T$& $16\gamma_{\phi}\chi_{\Xi}\geq(4\gamma_{\Xi}+\chi_{\phi})^{2}$& { Spin-rotation viscosity}\\
\hline
$\Xi^{\mu\lambda}=-u_\rho\nabla^{[\mu}\Omega^{\lambda]\rho}$& $\chi_{\Xi}=-2w_{19}/T$&$\chi_{\Xi}\geq 0$ &{ Spin-gyro-rotation viscosity}\\
\hline
$\mathcal{H}=\epsilon^{\rho\sigma\lambda\gamma}u_{\lambda}\partial_{\gamma}\Omega_{\rho\sigma}$& $\chi_{\mathcal{H}}=-w_2/T$& $\chi_{\mathcal{H}}\geq 0$& { Spin-gyro-pseudo bulk viscosity}\\
\hline
$\mathcal{I}^{\mu\rho}_{ S}=u_{\gamma}\epsilon^{\theta\lambda\gamma(\mu}\nabla^{\rho)}\Omega_{\theta\lambda}-\Delta^{\mu}_{\rho}\mathcal{H}/3$& $\chi_{\mathcal{I}}=w_{24}/(2T)$& $\chi_{\mathcal{I}}\geq 0$&{ Spin-gyro-pseudo shear viscosity}\\
\hline
\end{tabularx}
\caption{Summary of various transport coefficients for \( \delta S^{\mu\lambda\nu} \) in Eq.~\eqref{Spindissipativecurrent}.}
\label{Tablespin}
\end{table}

%
It is worth mentioning that most of the references cited in this section follow a 
similar approach in deriving the form of dissipative currents, assuming the validity of the
following local thermodynamic relations in spin hydrodynamics:
\be\label{tradtherm}
\begin{split}
    & Ts + \mu n = \rho + p - \frac{1}{2} \omega_{\mu\nu} S^{\mu\nu}\;, \\
    & \di p = s \, \di T + n \, \di \mu +\frac{1}{2} S^{\mu\nu} \di \omega_{\mu\nu}\;, 
\end{split}
\ee
where $\mu \equiv \zeta T$ is the chemical potential, $\rho$ the proper energy 
density, $p$ the pressure, $n$ the charge density, $S^{\mu\nu}\equiv u_{\lambda} \spt^{\lambda,\mu\nu}$ 
is the  ``spin density" and $\omega_{\mu\nu}= T \Omega_{\mu\nu}$.
Nevertheless, the differential relation in eq. \eqref{tradtherm} may not be fulfilled, as it was 
pointed out in Ref. \cite{Becattini:2023ouz}, so that using the equation \eqref{entropyrate} as 
a starting point appears to be a safer and more general method.\\

\section{Application to quantum kinetic theory}
\label{QKT}

Since the second law of thermodynamics is a macroscopic statement, any (first-order) microscopic theory has to comply with it. One such example is the formulation of spin hydrodynamics from quantum kinetic theory \cite{Weickgenannt:2022zxs, Weickgenannt:2022qvh, Wagner:2024fhf, Wagner:2024fry}.
Before being able to properly apply the results derived in the preceding sections, we have to clarify the power-counting used in quantum kinetic theory, which is referred to as the $\hbar$-gradient expansion and denotes an expansion around the classical limit. In particular, in the standard counting that treats spin-related quantities as small quantum corrections, one has $\hbar \partial_\lambda S^{\lambda\mu\nu}\sim T_A^{\mu\nu}\sim \mathcal{O}(\hbar^2)$. On the other hand, $j^\mu\sim T_S^{\mu\nu}\sim \mathcal{O}(1)$ are of zeroth order, with corrections emerging at second order. These corrections have not been computed in Refs. \cite{Weickgenannt:2022zxs, Weickgenannt:2022qvh, Wagner:2024fhf, Wagner:2024fry}, so we are a bit limited in our analysis.
We also have to revise the conditions given in Eqs. \eqref{matchingconditions-1} and \eqref{matchingconditions-2}. The reason is that, since in the aforementioned references the currents $\delta j^\mu$ and $\delta T_S^{\mu\nu}$ are only evaluated up to first order in $\hbar$, the matching conditions are $u_\mu \delta j^{(0)\mu}=0$ and $u_\mu \delta T_S^{(0)\mu\nu}=0$. Here, quantities marked with an index $(0)$ denote the contributions of zeroth order in quantum corrections.
The form of the currents is then
\begin{subequations}
\label{eqs:currents_aftermatching_kin}
\begin{align}
\delta j^\mu &= \mathcal{K}_\zeta^{(0)} \nabla^\mu\zeta  + \delta j^{(2)\mu}\;,\label{eq:delta_j_aftermatching_kin}\\
\delta T_S^{\mu\nu} &= \Delta^{\mu\nu}\zeta_b^{(0)} \theta + 2\eta^{(0)} \sigma^{\mu\nu} + \delta T_S^{(2)\mu\nu}\;,\label{eq:delta_TS_aftermatching_kin}\\
\delta T_A^{\mu\nu}&= 2u^{[\mu}\left(\kappa_h \mathcal{J}_h^{\nu]}-f_2 \mathcal{E}^{\nu]}+f_4 \mathcal{X}^{\nu]}-\kappa_\zeta \nabla^{\nu]}\zeta-\kappa_{\mathcal{G}} \mathcal{G}^{\nu]}\right) + \epsilon^{\mu\nu\alpha\beta} u_\alpha \left(-\gamma_\phi \mathcal{B}_\beta + \gamma_\Xi \mathcal{T}_\beta\right)\;,\label{eq:delta_TA_aftermatching_kin}\\
\delta S^{\lambda\mu\nu} &=\frac{1}{T}\Big[ - \frac12 u^{[\mu}\epsilon^{\nu]\lambda\alpha\beta} u_\alpha \left(-\chi_\phi \mathcal{B}_\beta  + \chi_\Xi \mathcal{T}_\beta\right) + \Delta^{\lambda[\mu}u^{\nu]} \left(k_1 T D\beta+ \chi_\theta \theta + \chi_\mathcal{Z} \mathcal{Z} + k_4 D\zeta\right)+ \chi_\sigma \sigma^{\lambda[\mu}u^{\nu]}+ \chi_\mathcal{F} \mathcal{F}_S^{\lambda[\mu}u^{\nu]} \nonumber\\
&\quad + \Delta^{\lambda[\mu}\left(\chi_h \mathcal{J}_h^{\nu]}-m_2 \mathcal{E}^{\nu]}+m_3 \mathcal{X}^{\nu]}+\chi_\zeta \nabla^{\nu]}\zeta+\chi_\mathcal{G} \mathcal{G}^{\nu]}\right) + \chi_\mathcal{H}\epsilon^{\lambda\mu\nu\alpha}u_\alpha  \mathcal{H} + \chi_\mathcal{I} \mathcal{I}_{S,\alpha}{}^{\lambda}\epsilon^{\mu\nu\alpha\rho}u_\rho\Big]\;.\label{eq:delta_S_aftermatching_kin}
\end{align}
\end{subequations}
The quantities $\delta j^{(2)\mu}$ and $\delta T_S^{(2)\mu\nu}$ are not further constrained. 
Compared to Eqs. \eqref{EMTsymmetricdissipativecurrent}--\eqref{Spindissipativecurrent}, there are a few differences, which are due to the different matching. In particular, in Eq. \eqref{eq:delta_TA_aftermatching_kin}, there are two extra coefficients, which in the matching discussed in the previous sections become $f_1=\kappa_h$ and $f_2=0$, respectively. Similarly, in Eq. \eqref{eq:delta_S_aftermatching_kin}, one would have $k_1=k_2=0$, $m_2=\chi_h$, and $m_3=0$.
The full set of inequalities that the coefficients have to fulfill in this case in order for the entropy production to be positive at any order in quantum corrections becomes hardly tractable. However, we can read off a set of necessary (though not sufficient) conditions,
\begin{align}
\label{eq:cond_limited}
    \mathcal{K}_\zeta^{(0)}&\geq 0 \;, \quad \zeta^{(0)}_b\geq 0 \;,\quad \eta^{(0)}\geq 0\;,\quad f_2 \geq 0\;,\quad \gamma_\phi \geq 0\;,\quad \chi_\Xi \geq 0\;,\nonumber\\
    \chi_\mathcal{Z} &\geq 0\;,\quad \chi_\mathcal{F}\geq 0\;,\quad \chi_\mathcal{G}\geq 0\;,\quad \chi_\mathcal{H} \geq 0\;,\quad \chi_\mathcal{I} \geq 0\;,
\end{align}
that is identical to a subset of entries in Tables \ref{Tablesymmetric}--\ref{Tablespin}, namely those that require certain coefficients to be nonnegative. Even though we do not have access to $\delta j^{(2)\mu}$ and $\delta T_S^{(2)\mu\nu}$, we can still check the inequalities \eqref{eq:cond_limited} explicitly.

For the comparison, we focus on the formulation derived in Ref. \cite{Wagner:2024fry}. Therein, the information on dissipation is included in the bulk viscous pressure $\Pi$, the diffusion current $n^\mu$, the shear-stress tensor $\pi^{\mu\nu}$, and a traceless rank-two tensor $\mathfrak{t}^{\mu\nu}$ appearing in the spin tensor. These quantities are dynamical and follow relaxation-type equations [detailed in Eq. (2) of Ref. \cite{Wagner:2024fry}]. To find their Navier-Stokes values, i.e., relate them algebraically to the first-
order quantities at our disposal, we only keep those terms in the respective equations of motion that are linear in the first-order tensors.\footnote{In the context of this work, we also treat the components of $\partial^\lambda\Omega^{\mu\nu}$ as being of first order when finding the Navier-Stokes limit of $\mathfrak{t}^{\mu\nu}$. Also, we identified $\mathcal{F}_S^{\mu\nu}\simeq\nabla^{\langle\mu}\kappa^{\nu\rangle}$.} In particular, keeping to the notation of Ref. \cite{Wagner:2024fry} and denoting potentially ambiguous quantities with a subscript ``kin'', we have
\begin{equation}
    \label{eq:NS_values}
    \Pi=-\zeta_\text{kin} \theta\;, \quad n^\mu = \kappa_\text{kin} \nabla^\mu \zeta \;,\quad \pi^{\mu\nu} = 2\eta_\text{kin} \sigma^{\mu\nu}\;,\quad \mathfrak{t}^{\mu\nu}= \frac{1}{T} \mathfrak{d} \sigma^{\mu\nu} +\ell_{\mathfrak{t}\kappa} \mathcal{F}_S^{\mu\nu}\;.
\end{equation}
Then, the form of the currents (in the Navier-Stokes limit) is given by
\begin{subequations}
\label{eqs:currents_kin}
\begin{align}
    \delta j^\mu_{\text{kin}}&= \kappa_{\text{kin}} \nabla^\mu \zeta+ \delta j^{(2)\mu}_{\text{kin}} \;,\\
    \delta T^{\mu\nu}_{S,\text{kin}}&= \zeta_{\text{kin}} \theta \Delta^{\mu\nu} +2 \eta_{\text{kin}} \sigma^{\mu\nu}+ \delta T_{S,\text{kin}}^{(2)\mu\nu}  \;,\\
    \delta T^{\mu\nu}_{A,\text{kin}}&= 2 u^{[\mu} \left(\frac{\Gamma^{(a)}}{2}\frac{1}{T}\mathcal{J}_h^{\nu]} - \frac{1}{T}\Gamma^{(\kappa)} \mathcal{E}^{\nu]}  + \Gamma^{(I)} \nabla^{\nu]}\zeta\right) -\frac{1}{2T} \Gamma^{(\omega)} \epsilon^{\mu\nu\alpha\beta} u_\alpha \mathcal{B}_\beta \;,\\
    \delta S^{\lambda\mu\nu}_{\text{kin}}&= \frac{2\sigma}{mT} \mathfrak{d}  \sigma^{\lambda[\mu} u^{\nu]}+\frac{2\sigma}{m}  \ell_{\mathfrak{t}\kappa}\mathcal{F}_S^{\lambda[\mu} u^{\nu]}\;,
\end{align}
\end{subequations}
with $\delta j^{(2)\mu}_{\text{kin}}$ and $\delta T_{S,\text{kin}}^{(2)\mu\nu}$ denoting the second-order contributions that have not been computed in Ref. \cite{Wagner:2024fry}.
Furthermore, $\sigma$ is the spin of the particle, and $I_{nq}$, $J_{nq}$ are thermodynamic integrals, which for Boltzmann statistics read
\begin{equation}
I_{nq}=J_{nq}=\frac{g e^{\alpha}}{(2q+1)!!} \frac{T^{n+2}}{2\pi^2} z^{n+2}\sum_{j=0}^{q+1} (-1)^j \binom{q+1}{j} \mathrm{Ki}_{2j-2-n}(z)\;, \label{eq:Inq_general_solution}
\end{equation}
with $z=m/T$ being the ratio of mass and temperature, $g=2\sigma + 1$ the degeneracy factor, and $\mathrm{Ki}_n(z)$ the Bickley function. 

Comparing Eqs. \eqref{eqs:currents_aftermatching_kin} and \eqref{eqs:currents_kin}, we can identify
\begin{align}
    \mathcal{K}^{(0)}_\zeta &\equiv \kappa_{\text{kin}} \;,\;\; \zeta^{(0)}_b \equiv \zeta_{\text{kin}}  \;,\;\; \eta^{(0)} \equiv \eta_{\text{kin}}\;,\nonumber\\
    \kappa_h&\equiv \frac{\Gamma^{(a)}}{2T} \;,\;\;f_2 \equiv \frac{\Gamma^{(\kappa)}}{T} \;,\;\; \kappa_\zeta \equiv - \Gamma^{(I)}\;, \;\; \gamma_\phi \equiv \frac{\Gamma^{(\omega)}}{2T} \;,\;\; \chi_\sigma \equiv  \frac{2\sigma}{m}\mathfrak{d} \;,\;\; \chi_\mathcal{F} \equiv \frac{2\sigma T}{m} \ell_{\mathfrak{t}\kappa}\;,\nonumber\\
    f_4 &=\kappa_\mathcal{G}=\gamma_\Xi=\chi_\phi =\chi_\Xi= k_1 = \chi_\theta = \chi_\mathcal{Z} = k_4 =
    \chi_h=m_2 =m_3=\chi_\zeta=\chi_\mathcal{G} = \chi_\mathcal{H} = \chi_\mathcal{I} \equiv 0\;.
\end{align} 
We emphasize that the fact that many coefficients vanish is a consequence of the simple truncation employed in Sec. IX of Ref. \cite{Wagner:2024fry}. When using a higher-order truncation, a lot of the coefficients will become nonzero, allowing for further nontrivial inequalities to be checked. These investigations are left for future work.
Considering the subset of coefficients that are nonzero, the inequalities \eqref{eq:cond_limited} become
\begin{equation}
\label{eq:ineqs_kin_1}
    \kappa_{\text{kin}}\geq 0\;,\quad \zeta_\text{kin} \geq 0\;,\quad \eta_{\text{kin}}\geq 0\;,\quad \Gamma^{(\kappa)}\geq 0\;,\quad \Gamma^{(\omega)}\geq 0\;,\quad \ell_{\mathfrak{t}\kappa} \geq 0\;.
\end{equation}
The first three inequalities are well-known from the case of standard hydrodynamics, while the following three appear only for nonzero spin. Note that the conditions $\Gamma^{(\kappa)}\geq 0$ and $\Gamma^{(\omega)}\geq 0$ can be related to the damping of the spin waves, i.e., the rate at which the reduced spin potential approaches the thermal vorticity \cite{Wagner:2024fhf}.

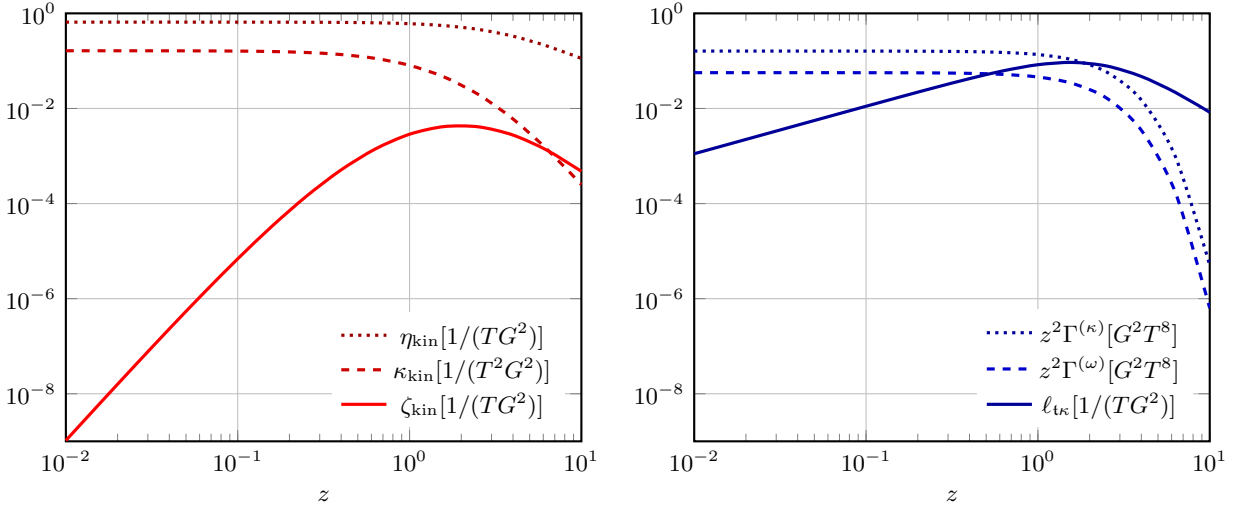
\begin{figure}
    \centering
     \begin{tikzpicture}
    \begin{axis}[name=plot1,enlargelimits=false,legend style={draw=none}, xmode=log, ymode=log,legend pos= south east, very thick, yticklabel style = {/pgf/number format/fixed}, grid=major, xmin=0.01, xmax=10,ymin=1e-9,ymax=1 ,axis line style = thick,
		xlabel=$z$,
		]
        \addplot [red!60!black,smooth, dotted] table[x index=0,y index=1,col sep=tab] {anc/TabEta.dat};
        \addlegendentry{$\eta_\text{kin}[1/(TG^2)]$}
        \addplot [red!80!black,smooth, dashed] table[x index=0,y index=1,col sep=tab] {anc/TabKappa.dat};
        \addlegendentry{$\kappa_\text{kin}[1/(T^2G^2)]$}
        \addplot [red!100!black,smooth] table[x index=0,y index=1,col sep=tab] {anc/TabZeta.dat};
        \addlegendentry{$\zeta_\text{kin}[1/(TG^2)]$}
		\end{axis}
        \begin{axis}[name=plot2, at={($(plot1.east)+(1.5cm,0)$)},anchor=west,enlargelimits=false,legend style={draw=none}, xmode=log, ymode=log,legend pos= south east, very thick, yticklabel style = {/pgf/number format/fixed}, grid=major, xmin=0.01, xmax=10,ymin=1e-9,ymax=1 ,axis line style = thick,
		xlabel=$z$,
		]
         \addplot [blue!60!black,smooth, dotted] table[x index=0,y index=1,col sep=tab] {anc/TabGammaKappa.dat};
    \addlegendentry{$z^2\Gamma^{(\kappa)}[G^2T^8]$}
        \addplot [blue!80!black,smooth, dashed] table[x index=0,y index=1,col sep=tab] {anc/TabGammaOmega.dat};
        \addlegendentry{$z^2\Gamma^{(\omega)}[G^2T^8]$}
        \addplot [blue!60!black,smooth] table[x index=0,y index=1,col sep=tab] {anc/TabltKappa.dat};
    \addlegendentry{$\ell_{\mathfrak{t}\kappa}[1/(TG^2)]$}
		\end{axis}
    \end{tikzpicture}
    \caption{Left: The transport coefficients that have to be semipositive also in hydrodynamics without spin. Right: The additional coefficients that have to be semipositive in spin hydrodynamics. Here, $G$ denotes the coupling strengh of the four-fermi interaction. Note that we multiplied the coefficients $\Gamma^{(\kappa)}$ and $\Gamma^{(\omega)}$ by $z^2$ to render the result finite.}
    \label{fig:check_ineq}
\end{figure}

While a formal proof for any interaction is outside the scope of this paper, we can check the inequalities \eqref{eq:ineqs_kin_1} for a specific microscopic interaction, which we choose to be a four-fermi one, $\mathcal{L}_\text{int}=G(\overline{\psi}\psi)^2$. As one can see in Fig. \ref{fig:check_ineq}, all coefficients are positive and the inequalities are indeed fulfilled. 

\section{Summary and outlook}
\label{Summary}
%
In summary, we have derived expressions for the first order dissipative corrections to the energy-momentum tensor, 
conserved vector current, and spin tensor incorporating the effects of spin potential gradients 
and finite chemical potential. The main results of this work are Eqs.~\eqref{EMTsymmetricdissipativecurrent}-\eqref{Spindissipativecurrent} and the corresponding transport coefficients, which are reported, named, 
and described in Tables~\ref{Tablesymmetric}-\ref{Tablespin}. In our derivation, we have expanded 
the tensor coefficients multiplying the gradients of the thermo-hydrodynamic fields from global thermodynamic 
equilibrium with vanishing thermal vorticity, thus taking advantage of rotational symmetry in the local rest 
frame.

With this foundation, one can formulate the evolution equations for the energy-momentum tensor, conserved vector 
current and spin tensor at first order in gradients, commonly referred to as the relativistic 
Navier-Stokes limit. However, the presence of 23 transport coefficients, which have to be either extracted from a suitable microscopic theory or fitted to data, may pose considerable challenges for 
future numerical studies, unless some of the gradients are neglected. Imposing the consequences of Onsager’s reciprocity theorem on our formalism may also relate some coefficients of different gradients.

It is important to note that the names assigned to each transport coefficient in Tables~\ref{Tablesymmetric}-\ref{Tablespin} 
are based on the best available physical understanding of their roles. A deeper comprehension of the underlying 
mechanisms would further refine our interpretation of each new coefficient.

\section*{Acknowledgements}
A.D. thanks the Department of Physics, University of Florence and INFN for the hospitality. A.D acknowleges 
technical discussions with Tomoki Goda and Radoslaw Ryblewski. A.D. acknowledges the financial support provided 
by the Polish National Agency for Academic Exchange NAWA  under the Programme STER–Internationalisation of 
doctoral schools, Project no.  PPI/STE/2020/1/00020 and the Polish National Science Centre Grants No.2018/30/E/ST2/00432.
D.W. acknowledges support by the project PRIN2022 Advanced Probes of the Quark Gluon Plasma funded by 
”Ministero dell’Università e della Ricerca”, the Deutsche Forschungsgemeinschaft (DFG, German Research Foundation)
through the CRC-TR 211 “Strong-interaction matter under extreme conditions” – project number 315477589 – TRR
211, and by the State of Hesse within the Research Cluster ELEMENTS (Project ID 500/10.006).

\appendix
%
\section{Decomposition of tensors into irreducible components under the rotation group}
\label{AppendixA}
%
In this section, we outline the method used to decompose the tensor coefficients $ H^{\mu\nu\rho\sigma}, 
K^{\mu\nu\rho}, \dots, W^{\mu\lambda\nu\rho\sigma\tau}$ in equation \eqref{Spingradientdecomposition} 
in terms of the rotationally invariant components reported in Sec.~\eqref{General Tensor Decomposition}. 
This method employs the irreducible representations of the rotation group \( \mathrm{SO}(3, \mathbb{R}) \). 

A four-vector $V^{\mu}$ can be decomposed into two irreducible components under the rotation group 
associated to the space orthogonal to $u$: a scalar, obtained by contracting $V$ with $u$ and a vector
with vanishing component along $u$. We will denote this decomposition of $V$ onto the irreducible 
representations $j=0$ and $j=1$ of \( \mathrm{SO}(3, \mathbb{R}) \) - that is scalar and the vector representations - by writing:
\be\label{vecdeco}
 V^{\mu}: (0\oplus 1) \to (u^{\mu}\oplus \Delta^{\mu}_{\alpha}) 
\ee
Note that we associate the four-vector $u^\mu$ to the scalar representation and the projector 
$\Delta^\mu_\alpha$ to the vector representation, where the extra index $\alpha$ signifies that the
vector $V^\mu$ is obtained by applying the projector $\Delta^{\mu}_{\alpha}$ to another vector.
Similarly, symmetric tensors $S^{\mu\nu}$ and antisymmetric tensors $A^{\mu\nu}$ can be expressed 
as follows:
\begin{align}\label{tensdeco}
    &S^{\mu\nu}: (0\oplus0\oplus1\oplus2) \to (u^{\mu}u^{\nu}\oplus \Delta^{\mu\nu}\oplus u^{(\mu}\Delta^{\nu)}_{\alpha}\oplus \Delta^{(\mu}_{\alpha}\Delta^{\nu)}_{\beta}-\frac{1}{3}\Delta_{\alpha\beta}\Delta^{\mu\nu}),\\ \nonumber 
    &A^{\mu\nu}: (1\oplus1) \to (u^{[\mu}\Delta^{\nu]\alpha}\oplus \epsilon^{\mu\nu\tau\alpha}u_{\tau}).
\end{align}
where the extra indices $\alpha$ and $\beta$ again signify that those expressions are projectors
(or proportional to projectors) to be applied to tensors. The $2$ in eq. \eqref{tensdeco} denotes 
the irreducible 5-dimensional representation pertaining to Euclidean symmetric traceless tensors. 
Note that the anti-symmetric tensor requires the use of the Levi-Civita tensor to generate 
an independent vector projector. 

By using the above decompositions as building blocks, we can decompose tensors of any rank 
into irreducible components under the rotation group. For example, let us consider the gradient 
of the reduced spin potential $\partial_{\rho}\Omega_{\lambda\nu}$, which is a rank-3 tensor antisymmetric 
in the last 2 indices. We can then write, by using the decompositions in \eqref{vecdeco} and 
\eqref{tensdeco}:
\be
    \partial_{\rho}\Omega_{\sigma\tau}: (0\oplus1)\otimes(1\oplus1)
    \to (u_{\rho}\oplus\Delta_{\rho\alpha})\otimes\left(\frac{1}{2}(u_{\sigma}\Delta_{\tau\beta}-u_{\tau}\Delta_{\sigma\beta})\oplus\epsilon_{\sigma\tau\gamma\beta}u^{\gamma}\right).\nonumber
\ee
where the first factor $(0 \oplus 1)$ refers to the gradient vector $\partial_\rho$ and the 
second, i.e., $(1\oplus1)$, to the reduced spin potential tensor $\Omega$. 
By expanding the above tensor product and taking into account the known rules for the decomposition
into irreducible representations, we get:
\begin{align}\label{listrep}
    &(0\otimes1)= 1 \to u_{\rho}u_{[\sigma}\Delta_{\tau]\beta}\nonumber\\
    &(0\otimes1)= 1 \to  u_{\rho}\epsilon_{\sigma\tau\gamma\beta}u^{\gamma}\nonumber\\
    &(1\otimes1) \to \Delta_{\rho}^{\alpha}u_{[\sigma}\Delta_{\tau]\beta}\nonumber\\
    &(1\otimes1)  \to \Delta^{\alpha}_{\rho}\epsilon_{\sigma\tau\gamma\beta}u^{\gamma}
\end{align}
We now can reduce the ($1 \otimes 1$) representation into $j=0,1,2$ and associate to each a
corresponding projector. In order to get the correct one for the scalar representation $j=0$,
one has to contract the indices $\alpha,\beta$ in the \eqref{listrep}; for the vector representation 
$j=1$ anti-symmetrize them by using a Levi-Civita projector $\epsilon^{\alpha\beta\lambda\,\delta}u_{\lambda}$ 
and for the tensor representation $j=2$ to symmetrize the indices $\alpha,\beta$ and remove 
the trace. Therefore:
\begin{align}
& (1\otimes1) \to \Delta_{\rho}^{\alpha}u_{[\sigma}\Delta_{\tau]\beta}=\begin{cases}
        0 \to u_{[\sigma}\Delta_{\tau]\rho}\nonumber\\
        1 \to \epsilon^{\alpha\beta\lambda\,\delta}u_{\lambda}\Delta_{\rho\alpha}u_{[\sigma}\Delta_{\tau]\beta}\\
        2 \to \Delta_{\rho(\alpha}u_{[\sigma}\Delta_{\tau]\beta)}-\frac{1}{3}
        \Delta_{\alpha\beta}u_{[\sigma}\Delta_{\tau]\rho}
    \end{cases}\nonumber\\
    &(1\otimes1) \to \Delta^{\alpha}_{\rho}\epsilon_{\sigma\tau\gamma\beta}u^{\gamma}=\begin{cases}
        0 \to \epsilon_{\sigma\tau\gamma\rho}u^{\gamma}\nonumber\\
        1 \to \epsilon^{\alpha\beta\lambda\,\delta}u_{\lambda}\Delta_{\rho\alpha}
        \epsilon_{\sigma\tau\gamma\beta}u^{\gamma}\nonumber\\
        2 \to \Delta_{\rho(\alpha}\epsilon_{\sigma\tau\gamma\beta)}u^{\gamma}-\frac{1}{3}\Delta_{\alpha\beta}\epsilon_{\sigma\tau\gamma\rho}u^{\rho}
    \end{cases}\nonumber
\end{align}
Altogether, $\partial_{\rho}\Omega_{\lambda\nu}$ can be decomposed into 8 irreducible components 
under rotation and written as:
\begin{align}  \partial_{\rho}\Omega_{\sigma\tau}=&c_1\mathcal{X}^{\gamma}u_{\rho}u_{[\sigma}\Delta_{\tau]\gamma}
+c_2\mathcal{Y}^{\gamma}u_{\rho}\epsilon_{\sigma\tau\lambda\gamma}u^{\lambda}
+c_3\mathcal{Z}u_{[\sigma}\Delta_{\tau]\rho}
+c_4\mathcal{T}^{\gamma}u_{[\sigma}\epsilon_{\rho\tau]\lambda\gamma}u^{\lambda}
+c_5\mathcal{F}^{\alpha\beta}\Delta_{\rho(\alpha}u_{[\sigma}\Delta_{\tau]\beta)}\nonumber\\
&+c_6\mathcal{H}\epsilon_{\sigma\tau\gamma\rho}u^{\gamma}
+c_7\mathcal{G}_{\gamma}\epsilon_{\rho}^{~~\beta\lambda\gamma}u_{\lambda}\epsilon_{\sigma\tau\alpha\beta}u^{\alpha}
+c_8\mathcal{I}^{\alpha\beta}\Delta_{\rho(\alpha}\epsilon_{\sigma\tau\gamma\beta)}u^{\gamma}.
\end{align}
where $c_i,\ i=1,2,\dots,8$, are arbitrary constants whose setting amounts to a rescaling of
the scalar, vector and tensor coefficients. Note that in the expression above we have
merged the trace term of the $j=2$ representations into the $j=0$ representations, without 
loss of generality. The above expression corresponds to the eq. (\ref{spinpotentialgradientdecomposition}) 
in the main text with a suitable choice of the constants $c_i,\ i=1,2,\dots,8$.

The same method can be applied to the decomposition of the tensor coefficients in the equation \eqref{Spingradientdecomposition}, with the remarkable simplification that we are only interested in
the scalar components. For instance, the tensor $H^{\mu\nu\rho\sigma}$ has the following decomposition:
\begin{align}\label{HdecompositionApp}
H^{\mu\nu\rho\sigma}=S^{\mu\nu}\otimes S^{\rho\sigma}: (0 \oplus 0 \oplus 1 \oplus 2) \otimes (0 \oplus 0 \oplus 1 \oplus 2).
\end{align}
Since:
\begin{align}
    0\,\otimes\,0 = 0 \qquad 0\, \otimes\, 1= 1 \qquad 1\, \otimes\, 1 = (0, 1, 2)  \qquad 1\, \otimes\, 2 = (1, 2, 3)
    \qquad 2\, \otimes\, 2 = (0, 1, 2,3,4)
\end{align}
according to the equation \eqref{HdecompositionApp}, the tensor $H^{\mu\nu\rho\sigma}$ can be 
decomposed as follows:
$$
H^{\mu\nu\rho\sigma}: 2( 0 \oplus 0 \oplus 1 \oplus 2) \oplus 1 \oplus 1 \oplus 0 \oplus 1 \oplus 2
\oplus 1 \oplus 2 \oplus 3 \oplus 2 \oplus 2 \oplus 1 \oplus 2 \oplus 3 \oplus 0 \oplus 1 \oplus 2 \oplus 3 \oplus 4
$$
As it can be inferred from the above expression, the tensor has six independent scalar components
and the projectors can be built by combining in all possible ways the elementary building blocks 
$u, \Delta$ and $\epsilon$. These are shown in Table~\ref{Table-G-I-O-F} in Sec.~\ref{General Tensor Decomposition}.

\bibliography{ref.bib}{}

@article{Wagner:2024fhf,
    author = "Wagner, David and Shokri, Masoud and Rischke, Dirk H.",
    title = "{On the damping of spin waves}",
    eprint = "2405.00533",
    archivePrefix = "arXiv",
    primaryClass = "nucl-th",
    month = "5",
    year = "2024"
}

@article{Hongo:2022izs,
    author = "Hongo, Masaru and Huang, Xu-Guang and Kaminski, Matthias and Stephanov, Mikhail and Yee, Ho-Ung",
    title = "{Spin relaxation rate for heavy quarks in weakly coupled QCD plasma}",
    eprint = "2201.12390",
    archivePrefix = "arXiv",
    primaryClass = "hep-th",
    reportNumber = "RIKEN-iTHEMS-Report-22",
    doi = "10.1007/JHEP08(2022)263",
    journal = "JHEP",
    volume = "08",
    pages = "263",
    year = "2022"
}

@article{Becattini:2024uha,
    author = "Becattini, Francesco and Buzzegoli, Matteo and Niida, Takafumi and Pu, Shi and Tang, Ai-Hong and Wang, Qun",
    title = "{Spin polarization in relativistic heavy-ion collisions}",
    eprint = "2402.04540",
    archivePrefix = "arXiv",
    primaryClass = "nucl-th",
    doi = "10.1142/S0218301324300066",
    journal = "Int. J. Mod. Phys. E",
    volume = "33",
    number = "06",
    pages = "2430006",
    year = "2024"
}

@article{Hidaka:2023oze,
    author = "Hidaka, Yoshimasa and Hongo, Masaru and Stephanov, Mikhail A. and Yee, Ho-Ung",
    title = "{Spin relaxation rate for baryons in a thermal pion gas}",
    eprint = "2312.08266",
    archivePrefix = "arXiv",
    primaryClass = "hep-ph",
    reportNumber = "RIKEN-iTHEMS-Report-23, KEK-TH-2565, J-PARC-TH-0294",
    doi = "10.1103/PhysRevC.109.054909",
    journal = "Phys. Rev. C",
    volume = "109",
    number = "5",
    pages = "054909",
    year = "2024"
}

@article{Wagner:2024fry,
    author = "Wagner, David",
    title = "{Resummed spin hydrodynamics from quantum kinetic theory}",
    eprint = "2409.07143",
    archivePrefix = "arXiv",
    primaryClass = "nucl-th",
    month = "9",
    year = "2024"
}

@article{Bhadury:2020puc,
    author = "Bhadury, Samapan and Florkowski, Wojciech and Jaiswal, Amaresh and Kumar, Avdhesh and Ryblewski, Radoslaw",
    title = "{Relativistic dissipative spin dynamics in the relaxation time approximation}",
    eprint = "2002.03937",
    archivePrefix = "arXiv",
    primaryClass = "hep-ph",
    doi = "10.1016/j.physletb.2021.136096",
    journal = "Phys. Lett. B",
    volume = "814",
    pages = "136096",
    year = "2021"
}

@article{STAR:2019erd,
    author = "Adam, Jaroslav and others",
    collaboration = "STAR",
    title = "{Polarization of $\Lambda$ ($\bar{\Lambda}$) hyperons along the beam direction in Au+Au collisions at $\sqrt{s_{_{NN}}}$ = 200 GeV}",
    eprint = "1905.11917",
    archivePrefix = "arXiv",
    primaryClass = "nucl-ex",
    doi = "10.1103/PhysRevLett.123.132301",
    journal = "Phys. Rev. Lett.",
    volume = "123",
    number = "13",
    pages = "132301",
    year = "2019"
}

@article{Florkowski:2017ruc,
    author = "Florkowski, Wojciech and Friman, Bengt and Jaiswal, Amaresh and Speranza, Enrico",
    title = "{Relativistic fluid dynamics with spin}",
    eprint = "1705.00587",
    archivePrefix = "arXiv",
    primaryClass = "nucl-th",
    doi = "10.1103/PhysRevC.97.041901",
    journal = "Phys. Rev. C",
    volume = "97",
    number = "4",
    pages = "041901",
    year = "2018"
}

@article{Gallegos:2021bzp,
    author = {Gallegos, A. D. and G\"ursoy, U. and Yarom, A.},
    title = "{Hydrodynamics of spin currents}",
    eprint = "2101.04759",
    archivePrefix = "arXiv",
    primaryClass = "hep-th",
    doi = "10.21468/SciPostPhys.11.2.041",
    journal = "SciPost Phys.",
    volume = "11",
    pages = "041",
    year = "2021"
}

@article{STAR:2017ckg,
    author = "Adamczyk, L. and others",
    collaboration = "STAR",
    title = "{Global $\Lambda$ hyperon polarization in nuclear collisions: evidence for the most vortical fluid}",
    eprint = "1701.06657",
    archivePrefix = "arXiv",
    primaryClass = "nucl-ex",
    doi = "10.1038/nature23004",
    journal = "Nature",
    volume = "548",
    pages = "62--65",
    year = "2017"
}

@article{Weickgenannt:2020aaf,
    author = "Weickgenannt, Nora and Speranza, Enrico and Sheng, Xin-li and Wang, Qun and Rischke, Dirk H.",
    title = "{Generating Spin Polarization from Vorticity through Nonlocal Collisions}",
    eprint = "2005.01506",
    archivePrefix = "arXiv",
    primaryClass = "hep-ph",
    doi = "10.1103/PhysRevLett.127.052301",
    journal = "Phys. Rev. Lett.",
    volume = "127",
    number = "5",
    pages = "052301",
    year = "2021"
}

@article{Hattori:2019lfp,
    author = "Hattori, Koichi and Hongo, Masaru and Huang, Xu-Guang and Matsuo, Mamoru and Taya, Hidetoshi",
    title = "{Fate of spin polarization in a relativistic fluid: An entropy-current analysis}",
    eprint = "1901.06615",
    archivePrefix = "arXiv",
    primaryClass = "hep-th",
    reportNumber = "RIKEN-iTHEMS-Report-19, YITP-19-15",
    doi = "10.1016/j.physletb.2019.05.040",
    journal = "Phys. Lett. B",
    volume = "795",
    pages = "100--106",
    year = "2019"
}

@article{Peng:2021ago,
    author = "Peng, Hao-Hao and Zhang, Jun-Jie and Sheng, Xin-Li and Wang, Qun",
    title = "{Ideal Spin Hydrodynamics from the Wigner Function Approach}",
    eprint = "2107.00448",
    archivePrefix = "arXiv",
    primaryClass = "hep-th",
    doi = "10.1088/0256-307X/38/11/116701",
    journal = "Chin. Phys. Lett.",
    volume = "38",
    number = "11",
    pages = "116701",
    year = "2021"
}

@article{Montenegro:2017rbu,
      author         = "Montenegro, David and Tinti, Leonardo and Torrieri,
                        Giorgio",
      title          = "{The ideal relativistic fluid limit for a medium with
                        polarization}",
      journal        = "Phys. Rev.",
      volume         = "D96",
      year           = "2017",
      number         = "5",
      pages          = "056012",
      doi            = "10.1103/PhysRevD.96.056012",
      eprint         = "1701.08263",
      archivePrefix  = "arXiv",
      primaryClass   = "hep-th",
      SLACcitation   = "%%CITATION = ARXIV:1701.08263;%%"
}

@article{She:2021lhe,
    author = "She, Duan and Huang, Anping and Hou, Defu and Liao, Jinfeng",
    title = "{Relativistic viscous hydrodynamics with angular momentum}",
    eprint = "2105.04060",
    archivePrefix = "arXiv",
    primaryClass = "nucl-th",
    doi = "10.1016/j.scib.2022.10.020",
    journal = "Sci. Bull.",
    volume = "67",
    pages = "2265--2268",
    year = "2022"
}

@article{Garbiso:2020puw,
    author = "Garbiso, Markus and Kaminski, Matthias",
    title = "{Hydrodynamics of simply spinning black holes \& hydrodynamics for spinning quantum fluids}",
    eprint = "2007.04345",
    archivePrefix = "arXiv",
    primaryClass = "hep-th",
    doi = "10.1007/JHEP12(2020)112",
    journal = "JHEP",
    volume = "12",
    pages = "112",
    year = "2020"
}

@article{Fu:2021pok,
    author = "Fu, Baochi and Liu, Shuai Y. F. and Pang, Longgang and Song, Huichao and Yin, Yi",
    title = "{Shear-Induced Spin Polarization in Heavy-Ion Collisions}",
    eprint = "2103.10403",
    archivePrefix = "arXiv",
    primaryClass = "hep-ph",
    doi = "10.1103/PhysRevLett.127.142301",
    journal = "Phys. Rev. Lett.",
    volume = "127",
    number = "14",
    pages = "142301",
    year = "2021"
}

@article{Daher:2022xon,
    author = "Daher, Asaad and Das, Arpan and Florkowski, Wojciech and Ryblewski, Radoslaw",
    title = "{Equivalence of canonical and phenomenological formulations of spin hydrodynamics}",
    eprint = "2202.12609",
    archivePrefix = "arXiv",
    primaryClass = "nucl-th",
    month = "2",
    year = "2022"
}

@article{Weickgenannt:2022qvh,
    author = "Weickgenannt, Nora and Wagner, David and Speranza, Enrico and Rischke, Dirk H.",
    title = "{Relativistic dissipative spin hydrodynamics from kinetic theory with a nonlocal collision term}",
    eprint = "2208.01955",
    archivePrefix = "arXiv",
    primaryClass = "nucl-th",
    doi = "10.1103/PhysRevD.106.L091901",
    journal = "Phys. Rev. D",
    volume = "106",
    number = "9",
    pages = "L091901",
    year = "2022"
}

@article{Daher:2022wzf,
    author = "Daher, Asaad and Das, Arpan and Ryblewski, Radoslaw",
    title = "{Stability studies of first-order spin-hydrodynamic frameworks}",
    eprint = "2209.10460",
    archivePrefix = "arXiv",
    primaryClass = "nucl-th",
    doi = "10.1103/PhysRevD.107.054043",
    journal = "Phys. Rev. D",
    volume = "107",
    number = "5",
    pages = "054043",
    year = "2023"
}

@article{Hu:2022azy,
    author = "Hu, Jin",
    title = "{Cross effects in spin hydrodynamics: A revisit from entropy analysis and statistical operator}",
    eprint = "2209.10979",
    archivePrefix = "arXiv",
    primaryClass = "nucl-th",
    month = "9",
    year = "2022"
}

@article{Drogosz:2024gzv,
    author = "Drogosz, Zbigniew and Florkowski, Wojciech and Hontarenko, Mykhailo",
    title = "{Hybrid approach to perfect and dissipative spin hydrodynamics}",
    eprint = "2408.03106",
    archivePrefix = "arXiv",
    primaryClass = "hep-ph",
    month = "8",
    year = "2024"
}

@article{Daher:2024bah,
    author = "Daher, Asaad and Florkowski, Wojciech and Ryblewski, Radoslaw and Taghinavaz, Farid",
    title = "{Stability and causality of rest frame modes in second-order spin hydrodynamics}",
    eprint = "2403.04711",
    archivePrefix = "arXiv",
    primaryClass = "hep-ph",
    doi = "10.1103/PhysRevD.109.114001",
    journal = "Phys. Rev. D",
    volume = "109",
    number = "11",
    pages = "114001",
    year = "2024"
}

@article{Ren:2024pur,
    author = "Ren, Xiang and Yang, Chen and Wang, Dong-Lin and Pu, Shi",
    title = "{Thermodynamic stability in relativistic viscous and spin hydrodynamics}",
    eprint = "2405.03105",
    archivePrefix = "arXiv",
    primaryClass = "nucl-th",
    doi = "10.1103/PhysRevD.110.034010",
    journal = "Phys. Rev. D",
    volume = "110",
    number = "3",
    pages = "034010",
    year = "2024"
}

@article{Tiwari:2024trl,
    author = "Tiwari, Abhishek and Patra, Binoy Krishna",
    title = "{Second-order spin hydrodynamics from Zubarev's nonequilibrium statistical operator formalism}",
    eprint = "2408.11514",
    archivePrefix = "arXiv",
    primaryClass = "hep-th",
    month = "8",
    year = "2024"
}

@article{Yang:2024duc,
    author = "Yang, Lixin and Yan, Li",
    title = "{Relativistic spin hydrodynamics revisited with general rotation by entropy-current analysis}",
    eprint = "2410.07583",
    archivePrefix = "arXiv",
    primaryClass = "nucl-th",
    month = "10",
    year = "2024"
}

@article{Dey:2024cwo,
    author = "Dey, Sourav and Das, Arpan",
    title = "{Kubo formula for spin hydrodynamics: spin chemical potential as leading order in gradient expansion}",
    eprint = "2410.04141",
    archivePrefix = "arXiv",
    primaryClass = "nucl-th",
    month = "10",
    year = "2024"
}

@article{Becattini:2023ouz,
    author = "Becattini, Francesco and Daher, Asaad and Sheng, Xin-Li",
    title = "{Entropy current and entropy production in relativistic spin hydrodynamics}",
    eprint = "2309.05789",
    archivePrefix = "arXiv",
    primaryClass = "nucl-th",
    doi = "10.1016/j.physletb.2024.138533",
    journal = "Phys. Lett. B",
    volume = "850",
    pages = "138533",
    year = "2024"
}

@article{Gallegos:2020otk,
    author = {Gallegos, A. D. and G\"ursoy, U.},
    title = "{Holographic spin liquids and Lovelock Chern-Simons gravity}",
    eprint = "2004.05148",
    archivePrefix = "arXiv",
    primaryClass = "hep-th",
    doi = "10.1007/JHEP11(2020)151",
    journal = "JHEP",
    volume = "11",
    pages = "151",
    year = "2020"
}

@article{Weickgenannt:2023bss,
    author = "Weickgenannt, Nora and Blaizot, Jean-Paul",
    title = "{Polarization dynamics from moment equations}",
    eprint = "2312.05917",
    archivePrefix = "arXiv",
    primaryClass = "hep-ph",
    doi = "10.1103/PhysRevD.109.056019",
    journal = "Phys. Rev. D",
    volume = "109",
    number = "5",
    pages = "056019",
    year = "2024"
}

@article{Becattini:2014yxa,
    author = "Becattini, F. and Bucciantini, L. and Grossi, E. and Tinti, L.",
    title = "{Local thermodynamical equilibrium and the beta frame for a quantum relativistic fluid}",
    eprint = "1403.6265",
    archivePrefix = "arXiv",
    primaryClass = "hep-th",
    doi = "10.1140/epjc/s10052-015-3384-y",
    journal = "Eur. Phys. J. C",
    volume = "75",
    number = "5",
    pages = "191",
    year = "2015"
}

@article{Becattini:2011zz,
    author = "Becattini, F.",
    title = "{Hydrodynamics of fluids with spin}",
    doi = "10.1134/S1547477111080036",
    journal = "Phys. Part. Nucl. Lett.",
    volume = "8",
    pages = "801--804",
    year = "2011"
}

@article{Gallegos:2022jow,
    author = "Gallegos, A. D. and Gursoy, U. and Yarom, A.",
    title = "{Hydrodynamics, spin currents and torsion}",
    eprint = "2203.05044",
    archivePrefix = "arXiv",
    primaryClass = "hep-th",
    month = "3",
    year = "2022"
}

@article{Weickgenannt:2022zxs,
    author = "Weickgenannt, Nora and Wagner, David and Speranza, Enrico and Rischke, Dirk H.",
    title = "{Relativistic second-order dissipative spin hydrodynamics from the method of moments}",
    eprint = "2203.04766",
    archivePrefix = "arXiv",
    primaryClass = "nucl-th",
    doi = "10.1103/PhysRevD.106.096014",
    journal = "Phys. Rev. D",
    volume = "106",
    number = "9",
    pages = "096014",
    year = "2022"
}

@article{Cartwright:2021qpp,
    author = "Cartwright, Casey and Amano, Markus Garbiso and Kaminski, Matthias and Noronha, Jorge and Speranza, Enrico",
    title = "{Convergence of hydrodynamics in rapidly spinning strongly coupled plasma}",
    eprint = "2112.10781",
    archivePrefix = "arXiv",
    primaryClass = "hep-th",
    month = "12",
    year = "2021"
}

@article{Fang:2025aig,
    author = "Fang, Shuo and Fukushima, Kenji and Pu, Shi and Wang, Dong-Lin",
    title = "{Relativistic spin hydrodynamics with antisymmetric spin tensors and an extension of the Bargmann-Michel-Telegdi equation}",
    eprint = "2506.20698",
    archivePrefix = "arXiv",
    primaryClass = "nucl-th",
    month = "6",
    year = "2025"
}

@article{Dey:2024crk,
    author = "Dey, Sourav and Bhadury, Samapan and Florkowski, Wojciech and Ryblewski, Radoslaw and Jaiswal, Amaresh",
    title = "{Energy-momentum correlators of fermions at finite temperature and density}",
    eprint = "2409.18912",
    archivePrefix = "arXiv",
    primaryClass = "hep-ph",
    doi = "10.1103/PhysRevD.110.116002",
    journal = "Phys. Rev. D",
    volume = "110",
    number = "11",
    pages = "116002",
    year = "2024"
}

@article{Buzzegoli:2025zud,
    author = "Buzzegoli, Matteo",
    title = "{Kubo formulas for spin polarization in dissipative relativistic spin hydrodynamics: a first-order gradient expansion approach}",
    eprint = "2502.15520",
    archivePrefix = "arXiv",
    primaryClass = "nucl-th",
    doi = "10.1007/JHEP07(2025)255",
    journal = "JHEP",
    volume = "07",
    pages = "255",
    year = "2025"
}

@article{Buzzegoli:2024mra,
    author = "Buzzegoli, M. and Palermo, A.",
    title = "{Emergent canonical spin tensor in the chiral symmetric hot QCD}",
    eprint = "2407.14345",
    archivePrefix = "arXiv",
    primaryClass = "hep-ph",
    month = "7",
    year = "2024"
}

@article{Florkowski:2024bfw,
    author = "Florkowski, Wojciech and Hontarenko, Mykhailo",
    title = "{Generalized Thermodynamic Relations for Perfect Spin Hydrodynamics}",
    eprint = "2405.03263",
    archivePrefix = "arXiv",
    primaryClass = "hep-ph",
    doi = "10.1103/PhysRevLett.134.082302",
    journal = "Phys. Rev. Lett.",
    volume = "134",
    number = "8",
    pages = "082302",
    year = "2025"
}

@article{Daher:2024ixz,
    author = "Daher, Asaad and Florkowski, Wojciech and Ryblewski, Radoslaw",
    title = "{Stability constraint for spin equation of state}",
    eprint = "2401.07608",
    archivePrefix = "arXiv",
    primaryClass = "hep-ph",
    doi = "10.1103/PhysRevD.110.034029",
    journal = "Phys. Rev. D",
    volume = "110",
    number = "3",
    pages = "034029",
    year = "2024"
}

@article{Bhadury:2025boe,
    author = "Bhadury, Samapan and Drogosz, Zbigniew and Florkowski, Wojciech and Kar, Sudip Kumar and Mykhaylova, Valeriya",
    title = "{Local equilibrium Wigner function for spin-1/2 particles}",
    eprint = "2505.02657",
    archivePrefix = "arXiv",
    primaryClass = "hep-ph",
    month = "5",
    year = "2025"
}

@article{Drogosz:2024lkx,
    author = "Drogosz, Zbigniew and Florkowski, Wojciech and {\L}ygan, Natalia and Ryblewski, Radoslaw",
    title = "{Boost-invariant spin hydrodynamics with spin feedback effects}",
    eprint = "2411.06154",
    archivePrefix = "arXiv",
    primaryClass = "hep-ph",
    doi = "10.1103/PhysRevC.111.024909",
    journal = "Phys. Rev. C",
    volume = "111",
    number = "2",
    pages = "024909",
    year = "2025"
}

@article{Kiamari:2023fbe,
    author = "Kiamari, M. and Sadooghi, N. and Jafari, M. Sedighi",
    title = "{Relativistic magnetohydrodynamics of a spinful and vortical fluid: Entropy current analysis}",
    eprint = "2310.01874",
    archivePrefix = "arXiv",
    primaryClass = "nucl-th",
    doi = "10.1103/PhysRevD.109.036024",
    journal = "Phys. Rev. D",
    volume = "109",
    number = "3",
    pages = "036024",
    year = "2024"
}

@article{Kumar:2023ojl,
    author = "Kumar, Avdhesh and Yang, Di-Lun and Gubler, Philipp",
    title = "{Spin alignment of vector mesons by second-order hydrodynamic gradients}",
    eprint = "2312.16900",
    archivePrefix = "arXiv",
    primaryClass = "nucl-th",
    doi = "10.1103/PhysRevD.109.054038",
    journal = "Phys. Rev. D",
    volume = "109",
    number = "5",
    pages = "054038",
    year = "2024"
}

@article{Liu:2021uhn,
    author = "Liu, Shuai Y. F. and Yin, Yi",
    title = "{Spin polarization induced by the hydrodynamic gradients}",
    eprint = "2103.09200",
    archivePrefix = "arXiv",
    primaryClass = "hep-ph",
    doi = "10.1007/JHEP07(2021)188",
    journal = "JHEP",
    volume = "07",
    pages = "188",
    year = "2021"
}

@article{Hongo:2021ona,
    author = "Hongo, Masaru and Huang, Xu-Guang and Kaminski, Matthias and Stephanov, Mikhail and Yee, Ho-Ung",
    title = "{Relativistic spin hydrodynamics with torsion and linear response theory for spin relaxation}",
    eprint = "2107.14231",
    archivePrefix = "arXiv",
    primaryClass = "hep-th",
    reportNumber = "RIKEN-iTHEMS-Report-21",
    doi = "10.1007/JHEP11(2021)150",
    journal = "JHEP",
    volume = "11",
    pages = "150",
    year = "2021"
}

@article{Bhadury:2021oat,
    author = "Bhadury, Samapan and Bhatt, Jitesh and Jaiswal, Amaresh and Kumar, Avdhesh",
    title = "{New developments in relativistic fluid dynamics with spin}",
    eprint = "2101.11964",
    archivePrefix = "arXiv",
    primaryClass = "hep-ph",
    doi = "10.1140/epjs/s11734-021-00020-4",
    journal = "Eur. Phys. J. ST",
    volume = "230",
    number = "3",
    pages = "655--672",
    year = "2021"
}

@article{Li:2020eon,
    author = "Li, Shiyong and Stephanov, Mikhail A. and Yee, Ho-Ung",
    title = "{Nondissipative Second-Order Transport, Spin, and Pseudogauge Transformations in Hydrodynamics}",
    eprint = "2011.12318",
    archivePrefix = "arXiv",
    primaryClass = "hep-th",
    doi = "10.1103/PhysRevLett.127.082302",
    journal = "Phys. Rev. Lett.",
    volume = "127",
    number = "8",
    pages = "082302",
    year = "2021"
}

@article{Montenegro:2018bcf,
    author = "Montenegro, David and Torrieri, Giorgio",
    title = "{Causality and dissipation in relativistic polarizable fluids}",
    eprint = "1807.02796",
    archivePrefix = "arXiv",
    primaryClass = "hep-th",
    doi = "10.1103/PhysRevD.100.056011",
    journal = "Phys. Rev. D",
    volume = "100",
    number = "5",
    pages = "056011",
    year = "2019"
}

@article{Shi:2020htn,
    author = "Shi, Shuzhe and Gale, Charles and Jeon, Sangyong",
    title = "{From chiral kinetic theory to relativistic viscous spin hydrodynamics}",
    eprint = "2008.08618",
    archivePrefix = "arXiv",
    primaryClass = "nucl-th",
    doi = "10.1103/PhysRevC.103.044906",
    journal = "Phys. Rev. C",
    volume = "103",
    number = "4",
    pages = "044906",
    year = "2021"
}

@article{Fukushima:2020ucl,
    author = "Fukushima, Kenji and Pu, Shi",
    title = "{Spin hydrodynamics and symmetric energy-momentum tensors \textendash{} A current induced by the spin vorticity \textendash{}}",
    eprint = "2010.01608",
    archivePrefix = "arXiv",
    primaryClass = "hep-th",
    doi = "10.1016/j.physletb.2021.136346",
    journal = "Phys. Lett. B",
    volume = "817",
    pages = "136346",
    year = "2021"
}

@article{Hattori:2019ahi,
      author         = "Hattori, Koichi and Hidaka, Yoshimasa and Yang, Di-Lun",
      title          = "{Axial Kinetic Theory and Spin Transport for Fermions
                        with Arbitrary Mass}",
      journal        = "Phys. Rev.",
      volume         = "D100",
      year           = "2019",
      number         = "9",
      pages          = "096011",
      doi            = "10.1103/PhysRevD.100.096011",
      eprint         = "1903.01653",
      archivePrefix  = "arXiv",
      primaryClass   = "hep-ph",
      reportNumber   = "RIKEN-QHP-408, RIKEN-iTHEMS-Report-19, YITP-19-14",
      SLACcitation   = "%%CITATION = ARXIV:1903.01653;%%"
}

@article{Becattini:2018duy,
      author         = "Becattini, F. and Florkowski, Wojciech and Speranza,
                        Enrico",
      title          = "{Spin tensor and its role in non-equilibrium
                        thermodynamics}",
      journal        = "Phys. Lett.",
      volume         = "B789",
      year           = "2019",
      pages          = "419-425",
      doi            = "10.1016/j.physletb.2018.12.016",
      eprint         = "1807.10994",
      archivePrefix  = "arXiv",
      primaryClass   = "hep-th",
      SLACcitation   = "%%CITATION = ARXIV:1807.10994;%%"
}

@article{Biswas:2023qsw,
    author = "Biswas, Rajesh and Daher, Asaad and Das, Arpan and Florkowski, Wojciech and Ryblewski, Radoslaw",
    title = "{Relativistic second-order spin hydrodynamics: An entropy-current analysis}",
    eprint = "2304.01009",
    archivePrefix = "arXiv",
    primaryClass = "nucl-th",
    doi = "10.1103/PhysRevD.108.014024",
    journal = "Phys. Rev. D",
    volume = "108",
    number = "1",
    pages = "014024",
    year = "2023"
}
\bibliographystyle{utphys}
\end{document}